\documentclass[aip,jcp,reprint,noshowkeys]{revtex4-1}
\usepackage{graphicx,dcolumn,bm,xcolor,microtype,multirow,amscd,amsmath,amssymb,amsfonts,physics,mhchem,pifont,wrapfig}

\usepackage{natbib}
\AtBeginDocument{\nocite{achemso-control}}

\definecolor{hughgreen}{HTML}{009900}
\usepackage{mathpazo,libertine}
\usepackage[normalem]{ulem}
\definecolor{darkgreen}{RGB}{0, 180, 0}
\newcommand{\titou}[1]{#1}

\usepackage{hyperref}
\hypersetup{
    colorlinks=true,
    linkcolor=blue,
    filecolor=blue,      
    urlcolor=blue,	
    citecolor=blue
}

\newcommand{\ra}{\rightarrow}
\newcommand{\T}[1]{#1^{\intercal}}\newcommand{\inv}[1]{#1^{-1}}
\newcommand{\PTtrans}[1]{#1^{\PT}}

\newcommand{\Ne}{n}
\newcommand{\Nbas}{N}
\newcommand{\Nea}{\Ne_{\alpha}}
\newcommand{\Neb}{\Ne_{\beta}}

\newcommand{\hH}{\Hat{H}}

\newcommand{\bA}{\mathbold{A}}
\newcommand{\bB}{\mathbold{B}}
\newcommand{\bC}{\mathbold{C}}
\newcommand{\bD}{\mathbold{D}}
\newcommand{\bF}{\mathbold{F}}
\newcommand{\bG}{\mathbold{G}}

\newcommand{\bI}{\mathbold{I}}

\newcommand{\bO}{\mathbold{0}}
\newcommand{\bP}{\mathbold{P}}
\newcommand{\bQ}{\mathbold{Q}}

\newcommand{\bS}{\mathbold{S}}
\newcommand{\bU}{\mathbold{U}}
\newcommand{\bX}{\mathbold{X}}
\newcommand{\bZ}{\mathbold{Z}}
\newcommand{\bV}{\mathbold{V}}
\newcommand{\bh}{\mathbold{h}}
\newcommand{\bSig}{\mathbold{\Sigma}}

\newcommand{\br}{\mathbold{r}}
\newcommand{\bx}{\mathbold{x}}
\newcommand{\by}{\mathbold{y}}

\newcommand{\be}{\mathbold{\epsilon}}
\newcommand{\blam}{\mathbold{\lambda}}
\newcommand{\bLam}{\mathbold{\Lambda}}

\newcommand{\bCa}{\bC_{\alpha}}
\newcommand{\bCb}{\bC_{\beta}}
\newcommand{\bc}{\mathbold{c}}

\newcommand{\bca}{\bc_{\alpha}}
\newcommand{\bcb}{\bc_{\beta}}

\newcommand{\cI}{\mathcal{I}}
\newcommand{\cP}{\mathcal{P}}

\newcommand{\sP}{P}
\newcommand{\cS}{\mathcal{S}}
\newcommand{\cT}{\mathcal{T}}
\newcommand{\cC}{\mathcal{C}}
\newcommand{\cK}{\mathcal{K}}
\newcommand{\PT}{\mathcal{PT}}
\newcommand{\CPT}{\mathcal{CPT}}
\newcommand{\cA}{\mathcal{A}}

\newcommand{\sigx}{\mathbold{\sigma}_x}
\newcommand{\sigy}{\mathbold{\sigma}_y}
\newcommand{\sigz}{\mathbold{\sigma}_z}

\newcommand{\mP}{\bP}
\newcommand{\mT}{\I \sigy}
\newcommand{\mPT}{\bU}
\newcommand{\mO}{\mathbold{0}}

\newcommand{\ka}{\ket*{\alpha}}
\newcommand{\kb}{\ket*{\beta}}

\newcommand{\phia}{\phi_{\alpha}}
\newcommand{\phib}{\phi_{\beta}}
\newcommand{\phii}{\phi_{i}}
\newcommand{\phiia}{\phi_{i \alpha}}
\newcommand{\phiib}{\phi_{i \beta}}
\newcommand{\tha}{\theta_\alpha}
\newcommand{\thb}{\theta_\beta}

\newcommand{\spP}{\hat{\pi}}
\newcommand{\spT}{\hat{\tau}}

\newcommand{\Wfn}{\Psi}
\newcommand{\WfnHF}{\Wfn_\text{HF}}
\newcommand{\WfnRHF}{\Wfn_\text{RHF}}
\newcommand{\WfnUHF}{\Wfn_\text{UHF}}
\newcommand{\phiUHF}{\phi_{\text{UHF}}}
\newcommand{\phiRHF}{\phi_{\text{RHF}}}
\newcommand{\InProd}[2]{\braket{#1}{#2}}
\newcommand{\InProdH}[2]{\InProd{#1}{#2}_\text{H}}
\newcommand{\InProdC}[2]{\InProd{#1}{#2}_\text{C}}
\newcommand{\melC}[3]{\mel{#1}{#2}{#3}_\text{C}}

\newcommand{\I}{\mathrm{i}}

\newcommand{\EHF}{E_\text{HF}}

\newcommand{\cMO}{C} 

\newcommand{\MO}[1]{\phi_{#1}}

\newcommand{\RCF}{R_\text{CF}} 
 
\newcommand{\chiL}{\chi_\text{L}}
\newcommand{\chiR}{\chi_\text{R}}
\newcommand{\sigg}{\sigma_\text{g}}
\newcommand{\sigu}{\sigma_\text{u}}

\newcommand{\gMin}{\hphantom{-}}
\newcommand{\gAlpha}{\vphantom{\alpha}}
\newcommand{\gBeta}{\vphantom{\beta}}

\newcommand{\opS}{\hat{s}}
\newcommand{\bopS}{\opS}
\newcommand{\opSx}{\opS_{x}}
\newcommand{\opSy}{\opS_{y}}
\newcommand{\opSz}{\opS_{z}}
\newcommand{\opSpm}{\opS_{\pm}}
\newcommand{\opSmp}{\opS_{\mp}}
\newcommand{\s}{s}
\newcommand{\ns}{n_{s}}
\newcommand{\ms}{m_{s}}
\newcommand{\cphase}{\gamma}
\newcommand{\ladphase}{\xi}

\newcommand{\cmark}{\color{darkgreen}{\text{\ding{51}}}}
\newcommand{\xmark}{\color{red}{\text{\ding{55}}}}

\newcommand{\updotl}{%
  \mathrel{\ooalign{\hfil$\vcenter{
    \hbox{$\scriptscriptstyle\bullet$}}$\hfil\cr$\upharpoonleft$\cr}
  }%
}
\newcommand{\updotr}{%
  \mathrel{\ooalign{\hfil$\vcenter{
    \hbox{$\scriptscriptstyle\bullet$}}$\hfil\cr$\upharpoonright$\cr}
  }%
}
\newcommand{\dwdotl}{%
  \mathrel{\ooalign{\hfil$\vcenter{
    \hbox{$\scriptscriptstyle\bullet$}}$\hfil\cr$\downharpoonleft$\cr}
  }%
}
\newcommand{\dwdotr}{%
  \mathrel{\ooalign{\hfil$\vcenter{
    \hbox{$\scriptscriptstyle\bullet$}}$\hfil\cr$\downharpoonright$\cr}
  }%
}
\newcommand{\vac}{%
  \mathrel{\ooalign{\hfil$\vcenter{
    \hbox{$\scriptscriptstyle\bullet$}}$\hfil\cr$ $\cr}
  }%
}

\newcommand{\uddot}{%
  \mathrel{\ooalign{\hfil$\vcenter{
    \hbox{$\scriptscriptstyle\bullet$}}$\hfil\cr$\upharpoonleft\downharpoonright$\cr}
  }%
}

\newcommand{\LCPQ}{Laboratoire de Chimie et Physique Quantiques (UMR 5626), Universit\'e de Toulouse, CNRS, UPS, France}
\newcommand{\UCAM}{Department of Chemistry, University of Cambridge, Lensfield Road, Cambridge, CB2 1EW, U.K.}

\begin{document}	

\title{$\PT$-Symmetry in Hartree--Fock Theory}

\author{Hugh G.~A.~\surname{Burton}}
\thanks{Corresponding author}
\email{hb407@cam.ac.uk}
\affiliation{\UCAM}
\author{Alex J.~W.~\surname{Thom}}
\affiliation{\UCAM}
\author{Pierre-Fran\c{c}ois \surname{Loos}}
\email{loos@irsamc.ups-tlse.fr}
\affiliation{\LCPQ}

\begin{abstract}
\begin{wrapfigure}[12]{o}[-1.2cm]{0.4\linewidth}
	\centering
	\includegraphics[width=\linewidth]{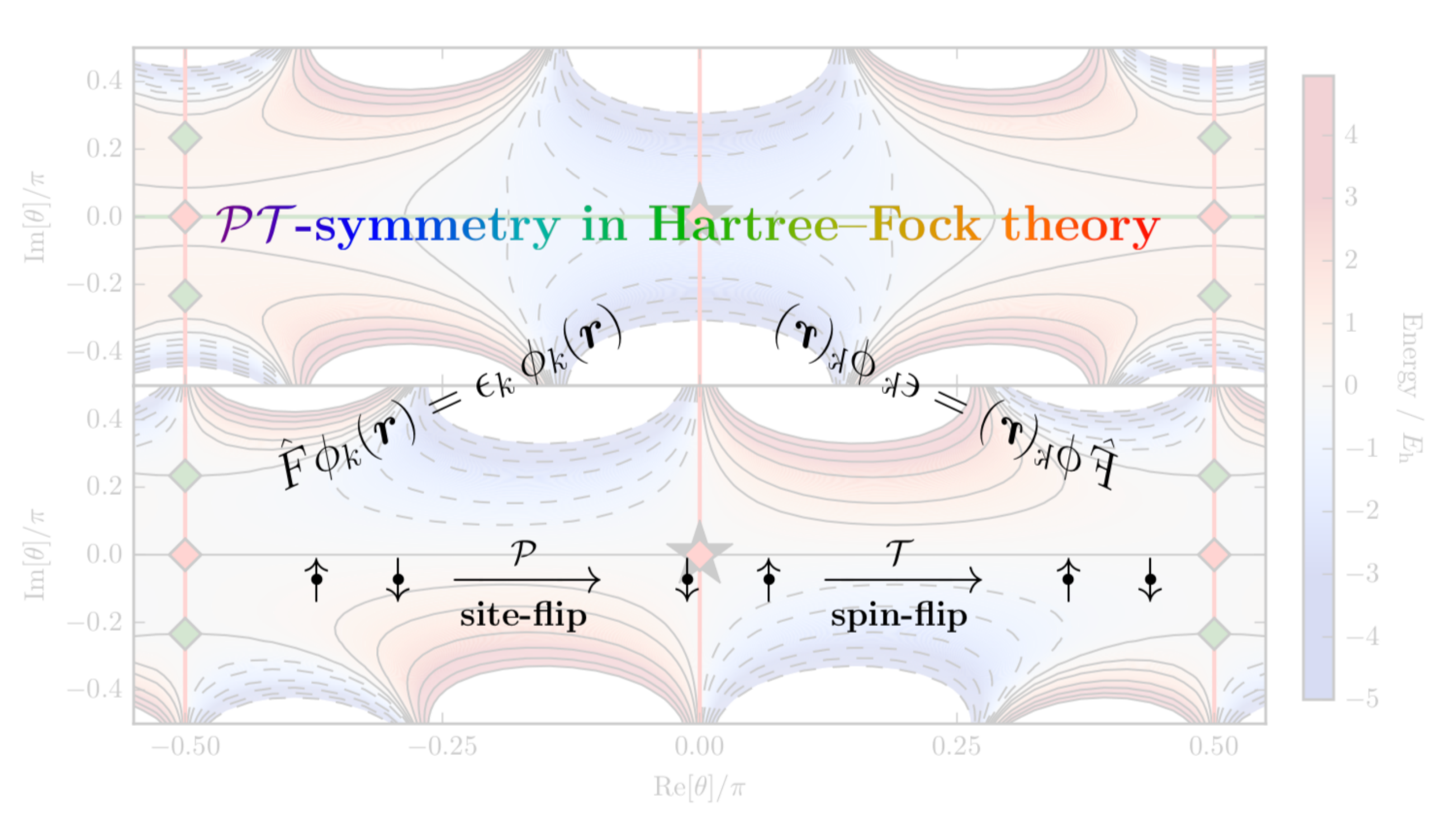}
\end{wrapfigure}
$\PT$-symmetry --- invariance with respect to combined space reflection $\cP$ and time reversal $\cT$ --- provides a weaker condition than (Dirac) Hermiticity for ensuring a real energy spectrum of a general non-Hermitian Hamiltonian.
$\PT$-symmetric Hamiltonians therefore form an intermediate class between Hermitian and non-Hermitian Hamiltonians.
In this work, we derive the conditions for $\PT$-symmetry in the context of electronic structure theory, and specifically, within the Hartree--Fock (HF) approximation.
We show that the HF orbitals are symmetric with respect to the $\PT$ operator \textit{if and only if} the effective Fock Hamiltonian is $\PT$-symmetric, and \textit{vice versa}.
By extension, if an optimal self-consistent solution is invariant under $\PT$, then its eigenvalues and corresponding HF energy must be real.
Moreover, we demonstrate how one can construct explicitly $\PT$-symmetric Slater determinants by forming $\PT$ doublets (i.e.~pairing each occupied orbital with its $\PT$-transformed analogue), allowing $\PT$-symmetry to be conserved throughout the self-consistent process.
Finally, considering the \ce{H2} molecule as an illustrative example, we observe $\PT$-symmetry in the HF energy landscape and find that the spatially symmetry-broken unrestricted HF wave functions (i.e.~diradical configurations) are $\PT$-symmetric, while the spatially symmetry-broken restricted HF wave functions (i.e.~ionic configurations) break $\PT$-symmetry. 
\end{abstract}

\maketitle

\section{Introduction}

Symmetry is an essential concept in quantum mechanics for describing properties that are invariant under particular transformations.
Physical observables, for example, must be totally symmetric under the group of symmetry operations corresponding to a quantum system, and the exact wave function must transform according to an irreducible representation of this group.
However, for approximate self-consistent methods such as Hartree--Fock (HF)\cite{SzaboBook} and Kohn--Sham density-functional theory (KS-DFT),\cite{ParrBook} occurrences of symmetry-breaking are pervasive and appear intimately linked to the breakdown of the single-determinant mean-field approximation in the presence of strong correlation.
From a chemical physicist's perspective, the archetypal example is the appearance of symmetry-broken HF solutions for internuclear distances beyond the so-called Coulson--Fischer point in \ce{H2} ($R > \RCF$),\cite{Coulson_1949} where the two (antiparallel) electrons localise on opposing nuclei with equal probability to form a spin-density wave. \cite{VignaleBook}

\begin{figure}
\includegraphics[width=\linewidth]{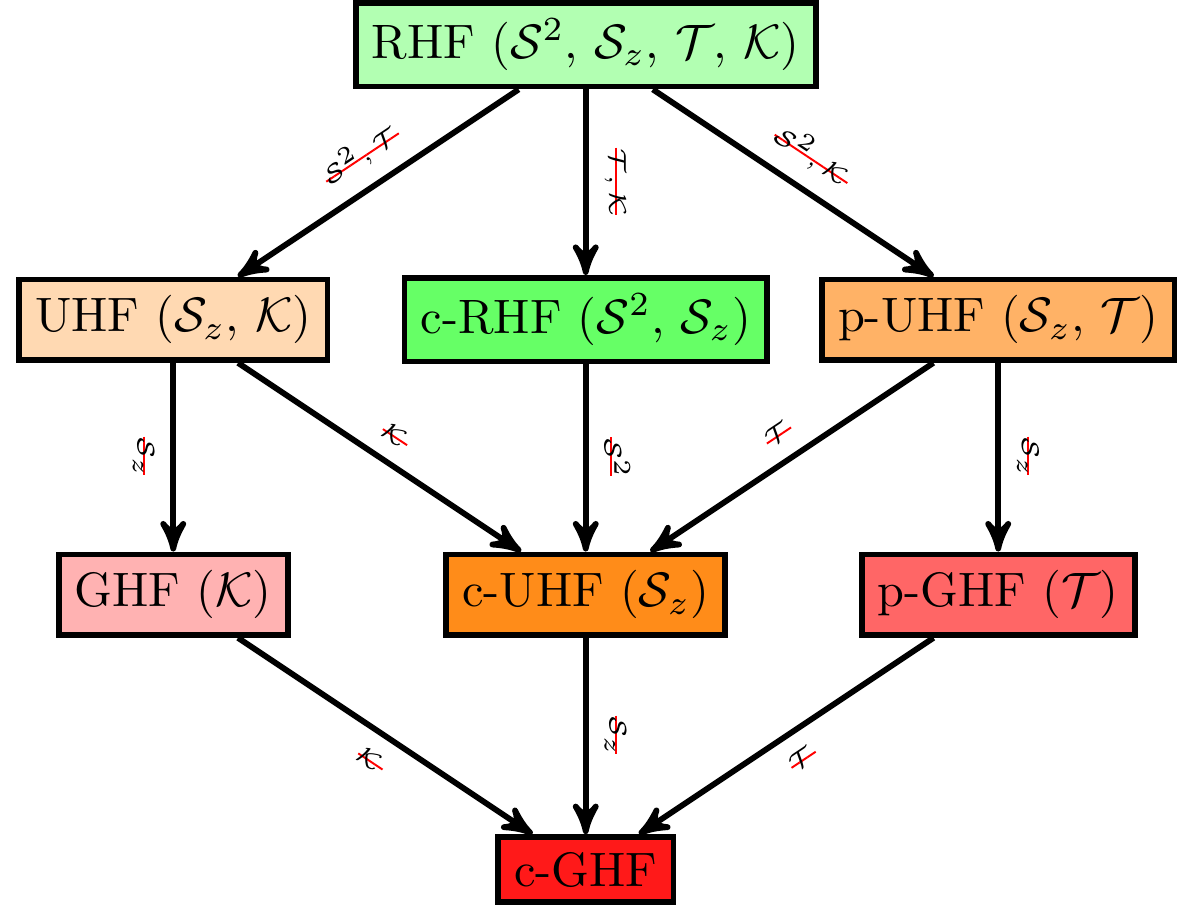}
\caption{The seven families of HF solutions, along with their definition according to Stuber and Paldus\cite{StuberPaldus} and the symmetries they conserve.
XHF, p-XHF and c-XHF stands for real, paired and complex XHF (where X $=$ R, U and G).
See main text for more details.}
\label{fig:summary_HF}
\end{figure}

Ensuring correct symmetries and good quantum numbers is critical in finite systems, especially since, when lost, their restoration is not always a straightforward task. \cite{Jimenez-Hoyos_2012, Cui_2013, Qiu_2017, Jake_2018}
However, applying symmetry ``constraints'' reduces flexibility, leading to the so-called \textit{symmetry dilemma} between variationally lower energies and good quantum numbers.\cite{Lykos_1963}
In general, approximate HF wave functions preserve only some of the symmetries of the exact wave function for finite systems (see Fig.~\ref{fig:summary_HF}).\cite{Fukutome_1981, StuberPaldus, Jimenez-Hoyos_2011}
The restricted HF (RHF) wave function, for example, forms an eigenfunction of the spin operators $\cS^2$ and $\cS_z$ by definition.
Additionally, restriction of the RHF wave function to real values ensures invariance with respect to time reversal $\cT$ and complex conjugation $\cK$.
By allowing the different spins to occupy different spatial orbitals in the unrestricted HF (UHF) approach, the wave function can break symmetry under $\cS^2$ but not $\cS_z$.
Constraining the UHF wave function to real values conserves $\cK$-symmetry, while the paired UHF (p-UHF) approach retains $\cT$-symmetry and the complex UHF (c-UHF) wave function can break both $\cK$- and $\cT$-symmetry.
The most flexible formulation, complex generalised HF (c-GHF), imposes none of these constraints, although paired (p-GHF) or real (GHF) variations maintain invariance with respect to $\cT$ or $\cK$ respectively.
All of these formalisms are independent of the point group symmetry (including the parity operator $\cP$), although spatial symmetry may be imposed separately on the HF wave function.

However, the HF approximation is not restricted to Hermitian approaches.
Holomorphic HF (h-HF) theory, for example, is formulated by analytically continuing real HF theory into the complex plane without introducing the complex conjugation of orbital coefficients.\cite{Hiscock_2014, Burton_2016, Burton_2018}
The result is a non-Hermitian Hamiltonian and an energy function that is complex analytic with respect to the orbital coefficients.
In addition, non-Hermitian HF approaches are extensively used to study unbound resonance phenomena where they occur in nature.\cite{MoiseyevBook}

Although initially intended as a method for extending symmetry-broken HF solutions beyond the Coulson--Fischer points at which they vanish,\cite{Hiscock_2014} h-HF theory also provides a more flexible framework for understanding the nature of multiple HF solutions in general.
For example, through the polynomial nature of the h-HF equations, a mathematically rigorous upper bound for the number of real RHF solutions can be derived for two-electron systems.\cite{Burton_2018}
Moreover, by scaling the electron-electron interactions using a complex parameter $\lambda$, h-HF theory reveals a deeper interconnected topology of multiple HF solutions across the complex plane.\cite{Burton_2019}
By slowly varying $\lambda$ in a similar (yet different) manner to an adiabatic connection in KS-DFT (without enforcing a density-fixed path),\cite{Seidl_2018} one can then ``morph'' a ground-state wave function into an excited-state wave function of a different symmetry via a stationary path of h-HF solutions, as we have recently demonstrated for a very simple model\cite{Seidl_2007, Loos_2009a, Loos_2009c, Loos_2015, Loos_2018a} in Ref.~\onlinecite{Burton_2019}.
In summary, h-HF theory provides a more general non-Hermitian framework with which the diverse properties of the HF approximation and its multiple solutions can be explored and understood.

In the present paper, we study a novel type of symmetry --- known as $\PT$-symmetry \cite{Bender_1998,Bender_1999,Bender_2002,Bender_2002a,Bender_2003,Bender_2004,Bender_2005,Bender_2006,Bender_2007,Bender_2007a,Bender_2008,Bender_2008a,Bender_2014,Bender_2015,Bender_2016,Bender_2017,Beygi_2018a,Liskow_1972,Peng_2014,Peng_2014a,BenderPTBook} --- in the context of electronic structure theory.
$\PT$-symmetry, i.e.~invariance with respect to combined space reflection $\cP$ and time reversal $\cT$, provides an alternative condition to (Dirac) Hermiticity which ensures real-valued energies even for complex, non-Hermitian Hamiltonians.\cite{Bender_2002}
Significantly, $\PT$-symmetric quantum mechanics allows the construction and study of many new types of Hamiltonians that would previously have been ignored.\cite{Bender_2005}
A Hermitian Hamiltonian, for example, can be analytically continued into the complex plane, becoming non-Hermitian in the process and exposing the fundamental topology of eigenstates.
Moreover, $\PT$-symmetric Hamiltonians can be considered as an intermediate class between Hermitian Hamiltonians commonly describing closed systems (i.e.~bound states) and non-Hermitian Hamiltonians which are peculiar to resonance phenomena (i.e.~open systems) where they naturally appear (see, for example, Ref.~\onlinecite{MoiseyevBook}).

Despite receiving significant attention across theoretical physics, \cite{BenderPTBook} to our knowledge $\PT$-symmetry remains relatively unexplored in electronic structure.
In the current work, we provide a first derivation of the conditions for $\PT$-symmetry in electronic structure, and specifically, within HF theory for closed systems.
By doing so, we hope to bridge the gap to $\PT$-symmetric physics, paving the way for future developments in electronic structure that exploit $\PT$-symmetry, for example novel wave function \textit{Ans\"{a}tze} or unusual approximate Hamiltonians.
Atomic units are used throughout.

\section{$\PT$-symmetric Hamiltonians}

\subsection{Spinless $\PT$-Symmetry}
To ensure a real energy spectrum and conservation of probability, it is commonly believed that a physically acceptable Hamiltonian $\hH$ must be Hermitian, i.e.~$\hH = \hH^\dag$, where $^\dag$ denotes the combination of complex conjugation ($^*$) and matrix transposition ($\T{}$).
Although the condition of Hermiticity is sufficient to ensure these properties, it is \textit{not} by any means necessary.
In particular, as elucidated by Bender and coworkers, \cite{Bender_1998} the family of $\PT$-symmetric Hamiltonians, \cite{Bender_1998, Bender_2016} defined such that $[\hH,\PT] = 0$ or $\hH = \PTtrans{\hH}$ where $\PTtrans{\hH} =(\PT) \hH \inv{(\PT)}$, provides a new more general class of Hamiltonians that allows for the possibility of non-Hermitian and complex Hamiltonians while retaining a physically sound quantum theory.\cite{Bender_2002}
Note that $[\cP,\cT] = 0$ but $\cP$ and/or $\cT$ may not commute with $\hH$.\cite{Bender_2007}

The textbook example of a $\PT$-symmetric Hamiltonian is \cite{BenderPTBook}
\begin{equation}
\label{eq:PT-H}
	\hH = p^2 + \I x^3
\end{equation}
which has been extensively studied by Bender and coworkers. \cite{Bender_1998, Bender_1999, Bender_2002, Bender_2002a, Bender_2003, Bender_2004, Bender_2005, Bender_2006, Bender_2007, Bender_2007a, Bender_2015, Bender_2016, Bender_2017}
\titou{From the standard action of $\cP$ and $\cT$, where
\begin{subequations}
\begin{align}
	\cP: & \quad p \ra -p, \quad x \ra -x,
	\\
	\cT: & \quad p \ra -p, \quad x \ra x, \quad \I \to -\I,
\end{align}
\end{subequations}
it is clear that the application of the combined space-time reflection $\PT$, where
\begin{align}
	\PT: & \quad p \ra p, \quad x \ra -x, \quad \I \to -\I,
\end{align}
leaves the Hamiltonian \eqref{eq:PT-H} unchanged.}
Moreover, although obviously complex, this Hamiltonian has a real, positive spectrum of eigenvalues!

Generalising the Hamiltonian \eqref{eq:PT-H} to the more general parametric family of $\PT$-symmetric Hamiltonians \cite{Bender_1998}
\begin{equation}
\label{eq:PT-H-eps}
	\hH = p^2 + x^2 (\I x)^\epsilon,	\quad	\epsilon \in \mathbb{R},
\end{equation}
one discovers a more complex structure.
It has been observed \cite{Bender_1998} and proved \cite{Dorey_2001} that, for $\epsilon \ge 0$ \titou{and a particular set of boundary conditions (the eigenfunctions must decay exponentially in well-defined sectors known as Stokes wedges \cite{BenderPTBook}),} the Hamiltonian \eqref{eq:PT-H-eps} has an entirely positive and real spectrum, while for $\epsilon < 0$, there are some complex eigenvalues which appear as complex conjugate pairs.
More specifically, in particular regions of parameter space, some eigenvalues coalesce and disappear by forming a pair of complex conjugate eigenvalues.
The region where some of the eigenvalues are complex is called the \textit{broken} $\PT$-symmetry region (i.e.~some of the eigenfunctions of $\hH$ are not simultaneously eigenfunctions of $\PT$), while the region where the entire spectrum is real is referred to as the \textit{unbroken} $\PT$-symmetry region.
Amazingly, these $\PT$-symmetry phase transitions have been observed experimentally in electronics, microwaves, mechanics, acoustics, atomic systems and optics, \cite{Bittner_2012, Chong_2011, Chtchelkatchev_2012, Doppler_2016, Guo_2009, Hang_2013, Liertzer_2012, Longhi_2010, Peng_2014, Peng_2014a, Regensburger_2012, Ruter_2010, Schindler_2011, Szameit_2011, Zhao_2010, Zheng_2013, Choi_2018, Goldzak_2018} and the parameter values where symmetry breaking occurs [$\epsilon = 0$ in the case of Hamiltonian \eqref{eq:PT-H-eps}] correspond to the appearance of exceptional points,\cite{Heiss_1990,Heiss_1991,Heiss_1999,Dorey_2009,Heiss_2012,Heiss_2016,Choi_2018,Lefebvre_2010,Liertzer_2012,Mailybaev_2005,Zhang_2018} the non-Hermitian analogues of conical intersections.\cite{Yarkony_1996}

\subsection{Electron $\PT$-Symmetry}

\titou{$\PT$-symmetric systems involving particles with non-zero spin, in our case electrons, are much less studied than their spinless counterparts.}
However, a number of studies have focused on this subject in recent years.\cite{Jones-Smith_2010, Cherbal_2012, Beygi_2018a, Beygi_2018b}
In what follows, we consider the spinor basis $\ket{\alpha} = \T{(1, 0)}$ and $\ket{\beta} = \T{(0, 1)}$.
A single-particle state is then represented by the column vector
\begin{equation}
\phi =
\begin{pmatrix}
    \phia \\
    \phib
\end{pmatrix},
\end{equation}
where $\phi_{\alpha}$ and $\phi_{\beta}$ are the $\alpha$ and $\beta$ components of $\phi$ respectively.
Note that, although a relativistic version of $\PT$-symmetric quantum mechanics can be formulated, \cite{Jones-Smith_2014} here we ignore effects such as spin-orbit coupling and consider only the non-relativistic limit.

The linear parity operator $\cP$ acts only on the spatial components and satisfies $\cP^2~=~\cI$, where $\cI$ is the identity operator.
Its action in the spinor basis can be represented by the block-diagonal matrix
\begin{equation}
	\cP \phi = 
	\begin{pmatrix}
	    \sP & 0  
	    \\
	    0 & \sP 
	\end{pmatrix} 
	\begin{pmatrix}
	    \phia
	    \\
	    \phib
	\end{pmatrix},
\label{eq:SpinorP}
\end{equation}
where $\sP$ represents the action of $\cP$ in the spatial basis.

In contrast, deriving the action of $\cT$ is a little more involved.
Fundamentally, $\cT$ is required to be an anti-linear operator, $\cT \I = - \I \cT$.\cite{QMModernDevelopment, WeinbergBook}
\titou{However, for systems containing particles with non-zero spin, the reversal of spin-angular momentum under the action of $\cT$ must also be included such that, for a given spin operator $\bopS$, we obtain $\cT \bopS = -\bopS \cT$.
Although a more detailed discussion on the nature of $\cT$ for particles of general spin is provided in Appendix~\ref{A:GeneralSpin}, here we focus on only the most relevant results.}

\titou{In the bosonic case, a basis can always be found in which $\cT$ is represented simply as $\cT = \cK$.\cite{Jones-Smith_2010}}
Here $\cK$ is the distributive anti-linear complex-conjugation operator which acts only to the right by convention \titou{and does not have a matrix representation}.\cite{QMModernDevelopment}
Applying $\cK$ in algebraic manipulations can lead to some non-intuitive results, and particular care must be exercised.
\titou{In contrast, the representation of $\cT$ for electrons (i.e. spin-$\frac{1}{2}$ particles) is given by $\cT = \I \sigy \cK$.\cite{Jones-Smith_2010}
To see why this must be the case,} consider expressing a spin operator $\bopS$ in a basis of the Pauli spin matrices $\qty( \sigx, \sigy, \sigz )$, where
\begin{align}
	\sigx & = 
	\begin{pmatrix}
		0 & 1 \\ 
		1 & 0
	\end{pmatrix},
	&
	\sigy & = 
	\begin{pmatrix}
		0 & -\I \\ 
		\I & \gMin 0 
	\end{pmatrix},
	&
	\sigz & = 
	\begin{pmatrix}
		1 & \gMin 0 \\ 
		0 & -1
	\end{pmatrix},
\end{align}
and $\bopS = \qty( \sigx, \sigy, \sigz )$.
Simply taking $\cT = \cK$ (as in the bosonic case) yields 
\begin{equation}
	\cT \bopS = \cK \qty( \sigx, \sigy, \sigz ) = \qty( \sigx, - \sigy, \sigz ) \cK,
\end{equation} 
which clearly does not give the desired outcome.
In contrast, taking the form $\cT = \I \sigy \cK$ yields\cite{Jones-Smith_2010}
\begin{equation}
	\cT \bopS
	= \I \sigy \cK \qty( \sigx, \sigy, \sigz )
	\\
	= - \qty( \sigx, \sigy, \sigz) \I \sigy \cK,
\end{equation}
therefore satisfying the correct behaviour $\cT \bopS = - \bopS \cT$.
Finally, consider also the behaviour of $\cT^2$, for which
\begin{equation}
	\cT^2	= \qty( \I \sigy \cK ) \qty( \I \sigy \cK )
			= \sigy \sigy \I^2 \cK^2 
			= - \sigy \sigy 
			= - \cI.
\end{equation}
Significantly, in fermionic systems, $\cT$ must be applied four times to return to the original state, leading to the action of time-reversal in fermionic systems being classified as odd.\cite{Jones-Smith_2010}

\titou{Overall, in the spinor basis,} the action of $\cT$ on $\phi$ can be represented by
\begin{equation}
\cT \phi = \mT \cK \phi \\
         = \begin{pmatrix}
                \gMin 0 & 1 \\
                - 1 & 0
            \end{pmatrix} \cK
            \begin{pmatrix}
                \phia
                \\
                \phib
            \end{pmatrix} \\
         = \begin{pmatrix}
                \gMin \phib^{*}
                \\
                - \phia^{*}
            \end{pmatrix}.
\label{eq:SpinorT}
\end{equation}
To find the representation of the combined $\PT$ operator, we simply combine the results of Eqs.~\eqref{eq:SpinorP} and \eqref{eq:SpinorT} to obtain
\begin{equation}
\begin{split}
\label{eq:PTActionRep}
	\PT \phi 
		& = \PTtrans{\phi}
		= \cP \mT \cK \phi
		\\
		& =
		\begin{pmatrix}
		    \gMin 0 & \sP 
		    \\
		    -\sP & 0
		\end{pmatrix} \cK
		\begin{pmatrix}
		    \phia
		    \\
		    \phib
		\end{pmatrix}
		= 
		\begin{pmatrix}
		    \gMin \sP \phib^{*}
		    \\
		    - \sP \phia^{*}
		\end{pmatrix}.
\end{split}
\end{equation}

\subsection{$\PT$-doublet}
\label{sec:PTdoublet}
A direct result of the odd character under $\cT$ is that it is impossible to find a single fermionic state $\phi$ that is invariant under the $\PT$ operator.
Instead, the closest analogue is a pair of states assembled into a $\PT$-doublet \cite{Jones-Smith_2010} of the form
\begin{equation}
\label{eq:PTDoublet}
    \begin{pmatrix}
	    \phi & -\PTtrans{\phi}
    \end{pmatrix}
    =
	\begin{pmatrix}
	    \phia^{\vphantom{*}} & -\sP \phib^{*} 
	    \\
	    \phib^{\vphantom{*}} & \gMin \sP \phia^{*}
	\end{pmatrix},
\end{equation}
where $\phi$ and $\PTtrans{\phi}$ are both eigenvectors of a $\PT$-symmetric Hamiltonian.
The action of $\PT$ on a $\PT$-doublet is then given by
\begin{equation}
\begin{split}
	\PT 
	\begin{pmatrix}
	    \phi & -\PTtrans{\phi}
    \end{pmatrix}
		& =		
            \begin{pmatrix}
		        \gMin 0 & \sP 
		        \\
		        - \sP & 0
		    \end{pmatrix} \cK
            \begin{pmatrix}
                \phia^{\vphantom{*}}		&	-\sP \phib^{*} \\
                \phib^{\vphantom{*}}		&	\gMin \sP \phia^{*}
            \end{pmatrix}      
         \\
         & =
            \begin{pmatrix}
                \gMin \sP \phib^{*}	&	\phia^{\vphantom{*}} 
                \\
                - \sP \phia^{*} 			&	\phib^{\vphantom{*}}
            \end{pmatrix}
          = 
          \begin{pmatrix} 
              \PTtrans{\phi} & \phi 
          \end{pmatrix},
\end{split}     
\end{equation} 
where the pair of eigenvectors have been simply swapped along with the introduction of a single minus sign.
We shall see later that the use of Slater determinants as antisymmetric many-electron wave functions enables strict $\PT$-invariance.
Note that invariance under $\PT$ implies that the energies of $\phi$ and $\PTtrans{\phi}$ are related by complex conjugation, while the additional assumption of unbroken $\PT$-symmetry implies that $\phi$ and $\PTtrans{\phi}$ must form degenerate pairs with real energies. \cite{Jones-Smith_2010}
Finally we note the inverse relationships $\inv{\cP} = \cP$ and $\inv{(\I \sigy)} = - \I \sigy = \T{ \I \sigy}$ which, in combination, yield $\inv{(\PT)} = - \I \sigy \cK \cP$. 

\section{$\PT$-Symmetry in Hartree--Fock}

\subsection{Hartree--Fock in practice}
\label{sec:HFpractice}

In the HF approximation, the wave function $\WfnHF$ for a system of $\Ne$ electrons is represented by a single Slater determinant constructed from a set of $\Ne$ occupied one-electron molecular orbitals $\MO{i}$ as
\begin{equation}
\label{eq:HFdet}
	\WfnHF = \cA \qty( \phi_{1} \phi_{2} \dots \phi_{\Ne} ),
\end{equation}
where $\cA$ is the anti-symmetrising operator.\cite{SzaboBook}
The single-particle orbitals $\phii$ are expanded in a finite-size direct product space of $\Nbas$ (one-electron) real spatial atomic orbital basis $\qty{\chi_1, \dots, \chi_{\Nbas}}$ and the spinor basis $\qty{\ka, \kb}$ as 
\begin{equation}
	\phii 
		= \sum_{\mu=1}^{\Nbas} \cMO^{\alpha}_{\mu i} \chi^{\gAlpha}_{\mu} \ka 
		+ \sum_{\mu=1}^{\Nbas} \cMO^{\beta}_{\mu i}  \chi^{\gBeta}_{\mu} \kb
		\\     
		= \phiia \ka + \phiib \kb,
\end{equation}
where $\phiia$ and $\phiib$ represent the $\alpha$ and $\beta$ components of $\phii$ respectively.
The coefficients $\cMO^{\alpha}_{\mu i}$ and $\cMO^{\beta}_{\mu i}$ are used to define the Slater determinant, and can be considered as components of a $\qty( 2\Nbas \times \Ne )$ matrix $\bC$ with the form 
\begin{equation}
\label{eq:CoeffRep}
	\bC = 
	\begin{pmatrix}
		\bCa	
		\\
		\bCb	
	\end{pmatrix},
\end{equation}
where $\bCa$ and $\bCb$ are $\qty( \Nbas \times \Ne )$ sub-matrices representing the expansions of $\phiia$ and $\phiib$.
  
In general, the atomic orbital basis set is not required to be orthogonal, although a real matrix $\bX$ can always be found such that $\T{\bX} \bS \bX = \bI$, where $\bS$ is the overlap matrix between atomic orbitals.
One particularly convenient choice is $\bX = \bS^{-1/2}$, but other choices are possible. \cite{SzaboBook}
Without loss of generality, we assume in the following that we are working in an orthogonal basis.

As an approximate wave function, $\WfnHF$ does not form an eigenfunction of the true electronic Hamiltonian $\hH$.
Instead, $\WfnHF$ is identified by optimising the HF energy $\EHF$ defined by the  expectation value for a given inner product $\InProd{\cdot}{\cdot}$ as
\begin{equation}
\label{eq:HFExpectedE}
	\EHF 
	= \frac{\InProd{\WfnHF}{\hH\ \WfnHF}}{\InProd{\WfnHF}{\WfnHF}}.
\end{equation}
The optimal set of HF molecular orbital coefficients $\bC$ are determined using a self-consistent procedure.
On each iteration $k$, an effective one-electron ``Fock'' Hamiltonian $\bF^{(k)}$ is constructed using the current occupied set of orbitals $\bC^{(k)}$, such that $\bF^{(k)} = \bh + \bD^{(k)} \bG$, where $\bh$ and $\bG$ are the one- and two-electron parts of the Fock matrix and $\bD^{(k)}$ is the density matrix at the $k$th iteration.
The new optimal molecular orbitals $\bC^{(k+1)}$ are then obtained by diagonalising $\bF^{(k)}$, i.e.~$\bF^{(k)}\bC^{(k+1)} = \bC^{(k+1)} \be^{(k+1)}$ where $\be$ is a diagonal matrix of the orbital energies, and the process is repeated until self-consistency is reached.\cite{SzaboBook}
At convergence we find $\bF \bD - \bD \bF = \bO$, demonstrating that, only at self-consistency, the Fock and density matrices commute.
(We drop the index $k$ for converged quantities.)
Note that $\bF$ is linear with respect to $\bD$, and that $\bh$ and $\bG$ are iteration independent and pre-computed at the start of the calculation.

\titou{Crucially, although the true $\Ne$-electron Hamiltonian $\hH$ is \emph{always} Hermitian, the process of dressing $\hH$ using the HF orbitals can lead to a non-Hermitian effective one-electron Hamiltonian.
In fact, the symmetry of $\bD^{(k)}$ and $\bF^{(k)}$ can depend of the specific choice of the inner product in Eq.~\eqref{eq:HFExpectedE}.}
For example, the most common choice is the Dirac Hermitian inner product $\InProdH{\bx}{\by} = \bx^{\dag} \by$, leading to Hermitian density $\bD^{(k)} = \bC^{(k)} (\bC^{(k)})^{\dag}$ and Fock matrices $\bF^{(k)} = (\bF^{(k)})^{\dag}$, and explicitly enforcing real energies.
Alternatively, the complex-symmetric inner product $\InProdC{\bx}{\by} = \T{\bx} \by$ requires complex-symmetric density  $\bD^{(k)} = \bC^{(k)} \T{(\bC^{(k)})}$ and Fock matrices $\bF^{(k)} = \T{(\bF^{(k)})}$, with energies that are complex in general.
\titou{In contrast to  complex-Hermitian HF, the complex-symmetric variant provides the unique analytic continuation of real HF for complex orbital coefficients.
This non-Hermitian formulation is used} in h-HF theory to ensure solutions exist over all geometries,\cite{Hiscock_2014, Burton_2016, Burton_2018, Burton_2019} and for describing resonance phenomena through non-Hermitian approaches.\cite{MoiseyevBook}

In what follows, we employ the complex-symmetric inner product $\InProd{.}{.} \equiv \InProdC{.}{.}$ to explore the conditions for $\PT$-symmetry under the HF approximation and understand under what circumstances $\EHF$ is real. 
\titou{In particular, we make use of the non-Hermitian h-HF formulation since this provides the natural mathematical extension of real h-HF for complex orbital coefficients.\cite{Burton_2018}}
We note that rigorous formulations of $\PT$-symmetric quantum mechanics introduce an additional linear operator $\cC$ and the $\CPT$ inner product to define a positive-definite inner product and ensure conservation of probability, although identifying $\cC$ is often non-trivial.\cite{Bender_2002a, BenderPTBook}
However, as an inherently approximate approach, HF theory requires only a well-defined inner product.
In our case, since the Fock matrix is explicitly complex-symmetric, its eigenvectors naturally form an orthonormal set under $\InProd{.}{.}_{\text{C}}$ without needing to introduce the $\CPT$ inner product.

\subsection{One-electron picture}
\label{sec:SingleParticlePic}

We turn now to the behaviour of the one-electron density, Fock matrices, and orbital energies under the $\PT$-operator.
First consider the relationship between the complex-symmetric density matrix $\bD = \bC \T{\bC}$ and the equivalent density matrix constructed using the $\PT$-transformed coefficients denoted $\PTtrans{\bC} = \PT \bC$.
The combined $\PT$ operator can be represented as the product $\PT = \mPT \cK$, where $\mPT$ is a $(2 \Nbas \times 2 \Nbas)$ real (linear) unitary matrix. \cite{Jones-Smith_2010}
Remembering that $\cK$ only acts on everything to the right, we subsequently find
\begin{equation}
\label{eq:PTCtoD}                             
\begin{split}
	\Big( \underbrace{\PTtrans{ \bC} }_{\mPT \cK \bC } \Big) \T{\qty(\PTtrans{\bC})} 
	& = \qty(\mPT \bC^{*}) \T{\qty(\mPT \bC^{*})}
	\\
	&= \underbrace{\mPT \cK}_{\PT} \bC \T{\bC} \underbrace{\cK\T{\mPT}}_{\inv{(\PT)}} 
	= \PTtrans{\bD}.
\end{split}
\end{equation}
where $\PTtrans{\bD} = (\PT) \bD \inv{(\PT)}$.
As a result, the density matrix constructed using the $\PT$-transformed coefficients is a $\PT$-similarity transformation of the density matrix constructed using the original set of coefficients.
Consequently, if a set of coefficients is $\PT$-symmetric, then the density matrix must be as well, and \textit{vice versa}.

Next consider the symmetry of the Fock matrix $\bF \qty[ \bD ]~=~\bh + \bD \bG$, which, due to its dependence on $\bD$, inherits the symmetry of the density used to construct it.
Assuming that $\bD$ is $\PT$-symmetric, i.e. $\bD = \PTtrans{\bD}$,  we find
\begin{equation}
\label{eq:DisPT_1}
    \bF[\bD] = \bh + \PTtrans{\bD} \bG 
	         = \bF[\PTtrans{\bD}].
\end{equation}
This result is trivial since $\bF$ is linear with respect to $\bD$.
Moreover, since the one- and two-electron parts of the Fock matrix are $\PT$-symmetric, i.e. $\bh = (\PT) \bh \inv{(\PT)}$ and $\bG = (\PT) \bG \inv{(\PT)}$,  we find
\begin{equation}
\label{eq:DisPT_2}
	(\PT) \bF[\bD] \inv{(\PT)} 
		= \bF[\PTtrans{\bD}].
\end{equation}
By equating Eqs.~\eqref{eq:DisPT_1} and \eqref{eq:DisPT_2} we see that, if $\bD$ is $\PT$-symmetric, then $\bF$ is also $\PT$-symmetric.
As a result, the symmetry of $\bF^{(k)}$ on a given iteration $k$ is dictated by the symmetry of the electron density from the current iteration $\bD^{(k)}$.
By extension, the symmetry of the new molecular orbitals $\bC^{(k+1)}$ is controlled by the symmetry of $\bF^{(k)}$ and, if one starts with a $\PT$-symmetric guess $\bD^{(0)}$, then $\PT$-symmetry can be conserved throughout the self-consistent process.
Furthermore, since $\bC^{(k+1)}$ is $\PT$-symmetric if and only if the effective Fock Hamiltonian $\bF^{(k)}$ is $\PT$-symmetric (and \textit{vice versa}), the existence of $\PT$-symmetry in HF can be identified by considering only the symmetry of the density itself.

Self-consistency of the HF equations requires the eigenvectors, which satisfy $\bF\qty[ \bD ] \bC = \bC \be$, to be equivalent to the coefficients used to build $\bD$ itself.
In other words, $\bF$ and $\bD$ commute and share the same set of eigenvectors.
Acting on the left with $\PT$ and exploiting the fact that $\inv{(\PT)} \PT = \cI$ yields
\begin{equation}
\begin{split}
	\PT \bF[\bD] \bC = \PT \bF[\bD] \inv{(\PT)} \PT \bC	& = \mPT \cK (\bC \be) 
	\\
	\implies \bF[\PTtrans{\bD}] (\PTtrans{\bC}) & = (\underbrace{\mPT \bC^{*}}_{\PTtrans{\bC}}) \be^*,
	\label{eq:PTFockMatrix}
\end{split}
\end{equation}
where we have used the result of Eq.~\eqref{eq:DisPT_2} and the property that $\cT$ is both anti-linear and distributive (see above) such that $\cK \bC \be = \bC^* \be^*$.
Combining with the result of Eq.~\eqref{eq:PTCtoD}, we can draw two conclusions.
Firstly, if a given set of orbital coefficients $\bC$ represents an optimised self-consistent HF solution with eigenvalues $\be$, then its $\PT$-transformed counterpart $\PTtrans{\bC}$ must also be a self-consistent solution with eigenvalues $\be^{*}$.
Secondly, and by extension, if an optimal self-consistent solution is invariant under $\PT$ (i.e.~$\bC = \PTtrans{\bC}$), then its eigenvalues must be real.

\subsection{Many-electron picture}
\label{sec:ManyParticlePic}

We turn now to the symmetry of the full HF Slater determinant $\WfnHF$ and its associated energy $\EHF$.
In the many-electron picture, the $\PT$-operator for an $\Ne$-electron system is given  as a product of one-electron operators,
\begin{equation}
	\PT = \bigotimes_{i=1}^{\Ne} \spP(i) \spT(i),
\end{equation}
where $\spP(i)$ and $\spT(i)$ are the parity and time-reversal operators acting only on the single-particle orbital occupied by electron $i$.
From the determinantal form of $\WfnHF$ [see Eq.~\eqref{eq:HFdet}], its symmetry under $\PT$ can be extracted as a product of its constituent orbitals symmetries.

Now, let us consider the relationship between the total HF energies of the two coefficient matrices $\bC$ and $\PTtrans{\bC}$ . 
Noting that $\T{\mPT} \mPT = \mPT \T{\mPT} = \bI$ and exploiting the invariance of the trace to cyclic permutations, i.e.~$\Tr( \bA \bB \bC) = \Tr( \bC \bA \bB)$, we find
\begin{equation}
\begin{split}
	\EHF[\bC] 
	& = \frac{1}{2} \Tr{\bD (\bh + \bF[\bD])} 
	\\
	& = \frac{1}{2} \Tr{\mPT \bD \T{\mPT} \mPT \qty(\bh + \bF[\bD]) \T{\mPT} }. 
\end{split}
\end{equation}
Note that $\cK^2 = \cI$ and, since $\cK$ acts only to the right, its application on the far right-hand side has no effect.
Therefore, by applying $\cK$ to both sides and inserting $\cK^2 = \cI$ in the middle, we find explicitly
\begin{equation}
	\EHF[\bC]^{*}
	= \cK \Tr{\mPT \bD \T{\mPT} \cK^{2} \mPT \qty(\bh +  \bF[\bD] ) \T{\mPT}} \cK
\end{equation}
Exploiting the distributive nature of $\cK$ over the matrix product within the trace, and since the reality of $\mPT$ provides $\cK \mPT = \mPT \cK$, we can migrate the $\cK$ operators to find
\begin{equation}
\begin{split}
   \EHF[\bC]^{*}
	& = \text{Tr} \Big\{ \underbrace{\mPT \cK}_{\PT} \bD \underbrace{\cK \T{\mPT}}_{\inv{(\PT)}} \underbrace{\mPT \cK}_{\PT} \qty(\bh + \bF[\bD] )\underbrace{\cK \T{\mPT}}_{\inv{(\PT)}} \Big\}  
	\\
	& = \Tr{\PTtrans{\bD} \qty(\bh + (\PT) \bF[\bD] \inv{(\PT)})}
	\\
	& = \Tr{\PTtrans{\bD} \qty(\bh + \bF[\PTtrans{\bD}])}  
	= \EHF[\PTtrans{\bC}],
\end{split}
\end{equation}
where we employ the result of Eq.~\eqref{eq:DisPT_2} and remember that $\bh$ is $\PT$-symmetric.
Overall we conclude that the respective HF energies corresponding to the coefficient matrices $\bC$ and $\PTtrans{\bC}$ are related by complex-conjugation.
Clearly by extension the HF energy of a $\PT$-symmetric set of orbital coefficients must be real.

\subsection{Hartree--Fock $\PT$-doublet}
\label{sec:PTdoubletHF}
To construct a set of occupied orbitals in the structure of a $\PT$-doublet [see Eq.~\eqref{eq:PTDoublet}], we require an explicit form of the matrix $\mPT$.
The linear parity operator $\cP$ acts only on the spatial basis and can be represented in the full direct product space by the Kronecker product $\bI_{2} \otimes \mP$, giving
\begin{equation}
	\cP \bC = 
	\begin{pmatrix}
	    \mP & \mO  
	    \\
	    \mO & \mP 
	\end{pmatrix} 
	\begin{pmatrix}
	    \bCa
	    \\
	    \bCb
	\end{pmatrix},
\end{equation}
where $\mP$ is a real $( \Nbas \times \Nbas )$ matrix representation of $\cP$ in the spatial basis, satisfying $\mP^2 = \bI_{\Nbas}$.
($\bI_{\Nbas}$ denotes the identity matrix of size $\Nbas$.)
As a result, the combined $\PT$ operator can be represented by the $(2 \Nbas \times 2 \Nbas)$ matrix constructed from the Kronecker product $\mPT = (\mT) \otimes \mP$, such that
\begin{equation}
\label{eq:PTActionRep2}
	\PT \bC 
		= \mPT \cK \bC
		\\
		= 
		\begin{pmatrix}
		    \hphantom{-}\mO & \mP 
		    \\
		    -\mP & \mO
		\end{pmatrix} \cK
		\begin{pmatrix}
		    \bCa
		    \\
		    \bCb
		\end{pmatrix}
		\\
		= 
		\begin{pmatrix}
		    \gMin \mP \bCb^{*}
		    \\
		    - \mP \bCa^{*}
		\end{pmatrix}.
\end{equation}

In the coefficient matrix representation [see Eq.~\eqref{eq:CoeffRep}], a $\PT$-doublet can then be constructed by pairing each occupied orbital with its $\PT$-transformed analogue, giving
\begin{equation}
\label{eq:PTDoubletC}
\bC = 
  \begin{pmatrix}
     \bc & -\PT \bc
  \end{pmatrix}
  =
  \begin{pmatrix}
     \bca^{\vphantom{*}} & - \mP \bcb^{*}
     \\
     \bcb^{\vphantom{*}} & \gMin \mP \bca^{*}
  \end{pmatrix},
\end{equation}
where $\bc$ and $-\PT \bc$ form $\qty( 2\Nbas \times \Ne/2 )$ sub-matrices representing the paired orbitals of the $\PT$-doublet.
The action of $\PT$ on a $\PT$-doublet is then 
\begin{equation}
\label{eq:PT1}
\begin{split}
	\PT 
	& \begin{pmatrix}
		\bc & -\PT \bc
	\end{pmatrix}
    \\	
	& = \mPT \cK
		\begin{pmatrix}
			\bca^{\vphantom{*}} & - \mP \bcb^{*}
			\\
			\bcb^{\vphantom{*}} & \gMin \mP \bca^{*}
		\end{pmatrix}      
	= \begin{pmatrix}
			\gMin \mP \bcb^{*}	&	\bca^{\vphantom{*}} 
			\\
			- \mP \bca^{*} 	&	\bcb^{\vphantom{*}}
		\end{pmatrix}
	\\	
	& = 
	\begin{pmatrix}
		\PT \bc & \bc
	\end{pmatrix}
	=
	\begin{pmatrix}
		\bc & -\PT \bc
	\end{pmatrix},
\end{split}     
\end{equation}
where the last line exploits the anti-symmetry of a determinantal wave function under the permutation of two columns in $\bC$.
Moreover, since the many-electron representation of $(\PT)^2$ is given by
\begin{equation}
\label{eq:PT2}
\begin{split}
	(\PT)^2 
	& = \bigotimes_{i=1}^{\Ne} \spP(i)^2   \spT(i)^2  = (-1)^\Ne \cI,
\end{split}
\end{equation}
we see that it is only possible to define a $\PT$-symmetric state in systems with $m_s = 0$ (i.e.~$\Nea = \Neb = \Ne/2$) where the occupied orbitals are paired in the structure of a $\PT$-doublet of the form given by Eq.~\eqref{eq:PTDoubletC}.

The behaviour of a $\PT$-doublet can be illustrated by considering a simple two-electron Slater determinant constructed from the orbitals $(\phi, -\PTtrans{\phi})$
\begin{equation}
\begin{split}
	\Wfn 
	& = \frac{1}{\sqrt{2}}
	\begin{vmatrix}
		\phi(1)	&	-\PTtrans{\phi}(1)	\\
		\phi(2)	&	-\PTtrans{\phi}(2)	\\
	\end{vmatrix}
	\\
	& = \frac{- \phi(1) \PTtrans{\phi}(2) + \PTtrans{\phi}(1) \phi(2)}{\sqrt{2}}.
\end{split}
\end{equation}
Thanks to the linearity and antisymmetry properties of determinants, Eq.~\eqref{eq:PT1} immediately yields
\begin{equation}
	\PT\,\Wfn = 
	\frac{1}{\sqrt{2}}
	\begin{vmatrix}
		\PTtrans \phi(1)	&	\phi(1)	\\
		\PTtrans \phi(2)	&	\phi(2)	\\
	\end{vmatrix}
	= \Wfn.
\end{equation}

\section{Example of \ce{H2}}

We now turn our attention to the didactic example of the \ce{H2} molecule in a minimal molecular orbital (orthogonal) basis
\begin{subequations}
\begin{align}
	\sigg & = (\chiL + \chiR)/\sqrt{1+2S},
	\\
	\sigu & = (\chiL - \chiR)/\sqrt{1-2S},
\end{align}
\end{subequations}
where $\chiL$ and $\chiR$ are the left and right atomic orbitals and $S = \InProd{\chiL}{\chiR}$ defines their overlap.
Without loss of generality, this paradigmatic two-electron system can be considered as a one-dimensional system, and the spatial representation of the parity operator in the $(\sigg,\sigu)$ basis is given by
\begin{equation}
	\mP = 
	\begin{pmatrix}
		1	&	\hphantom{-}0 
		\\
		0	&	-1
	\end{pmatrix}.
\end{equation}
In the following, all calculations are performed with the STO-3G (minimal) atomic basis.
For the sake of simplicity we focus on the $m_s = 0$ spin manifold.

\subsection{Real Orbital Coefficients}
\label{subsec:RealCase}

\begin{figure}[th]
	\includegraphics[width=\linewidth]{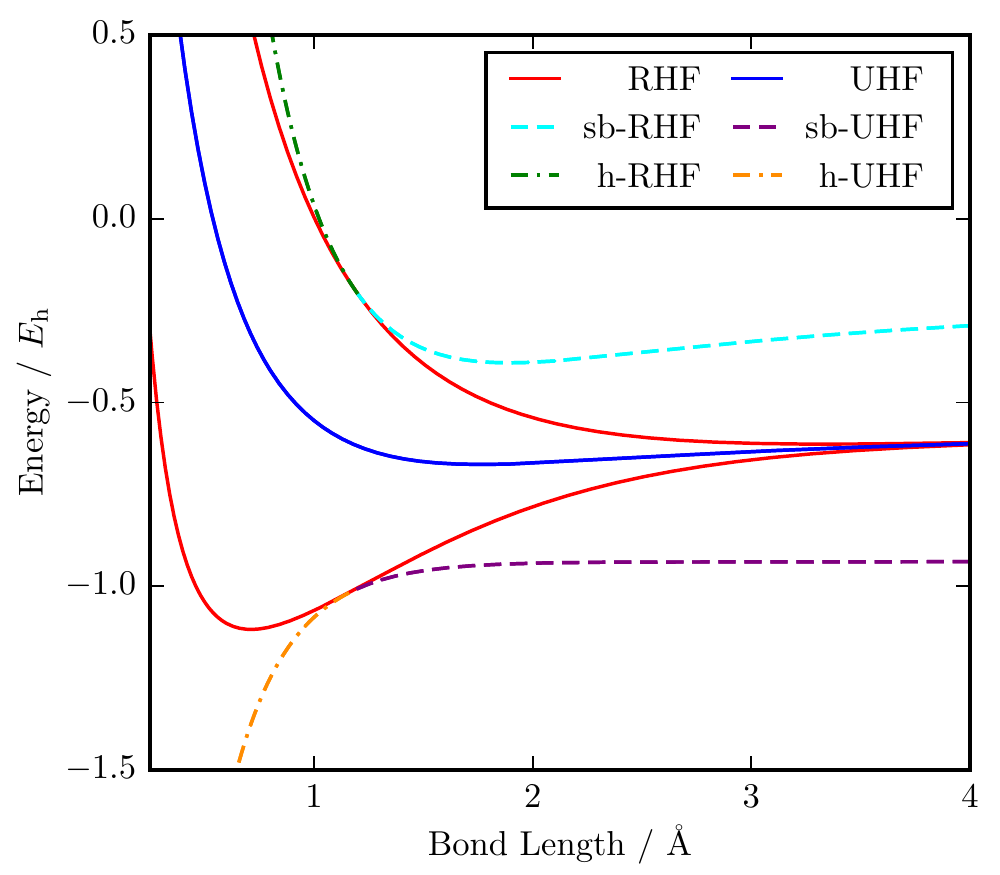}
	\caption{h-HF energy for the multiple solutions of \ce{H2}.
    The \titou{spatially} symmetry-pure solutions, i.e., the lowest RHF, UHF and highest RHF solutions correspond to the $\sigg^2$, $\sigg\sigu$ and $\sigu^2$ configurations, respectively.
    In the dissociation limit the \titou{spatially} symmetry-broken UHF (sb-UHF) states and \titou{spatially} symmetry-broken RHF (sb-RHF) states correspond to diradical configurations (\ce{^{$\upharpoonleft$}H-H^{$\downharpoonright$}} and \ce{^{$\downharpoonleft$}H-H^{$\upharpoonright$}}) and ionic configurations (\ce{H^{$+$}-H^{$-$}} and \ce{H^{$-$}-H^{$+$}}), respectively. 
	At shorter bond lengths, the sb-RHF and sb-UHF states coalesce with the spatially symmetry-pure RHF solutions and extend into the complex plane as h-RHF and h-UHF states, respectively.
	The holomorphic energy, $\melC{\WfnHF}{\hH}{\WfnHF}$, of the h-RHF and h-UHF solutions, however, remains real.}
	\label{fig:H2BindingCurve}
\end{figure}

\begin{figure}[tb]
	\includegraphics[width=\linewidth]{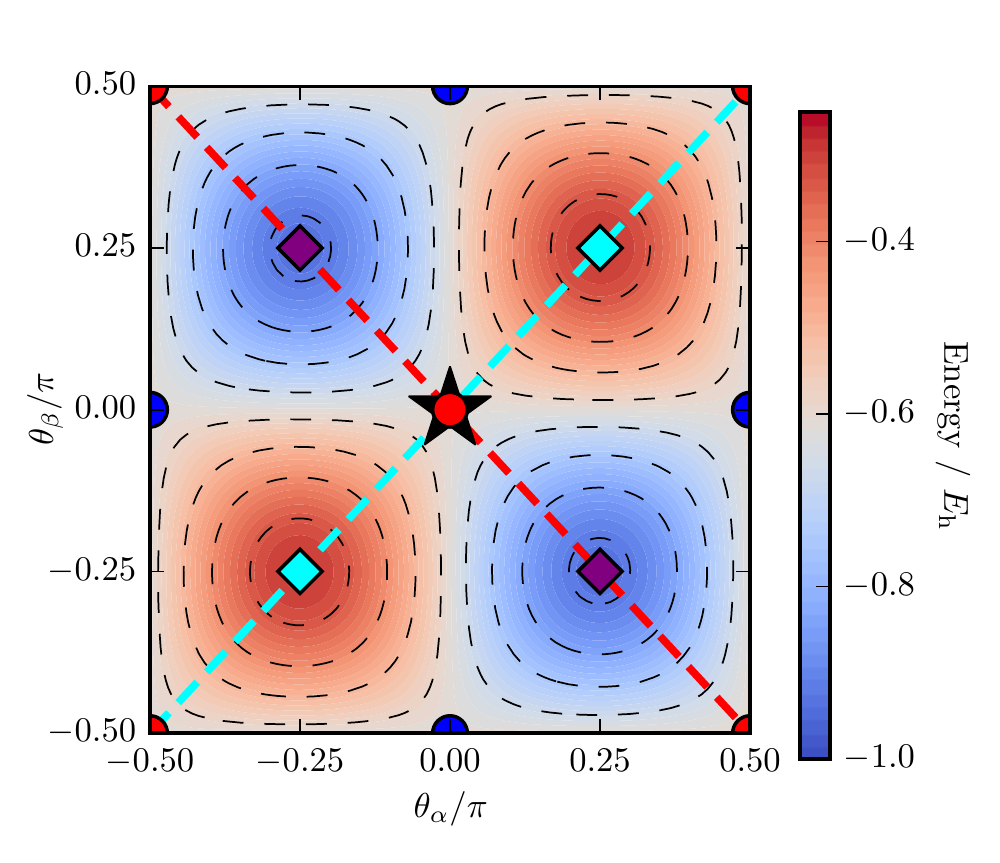}
	\caption{
	The UHF energy of \ce{H2} at $4$~\r{A} bond length as a function of $\tha$ and $\thb$ showing spin-flip symmetry (dashed cyan line) and $\PT$-symmetry (dashed red line).
	The parity (site-flip) operation corresponds to the mapping $(\tha, \thb) \ra (-\tha, -\thb)$, as indicated by inversion through the black star.
	Spatial symmetry-pure stationary points are indicated by red (RHF) and blue (UHF) circles, while stationary points breaking spatial symmetry are illustrated by cyan (sb-RHF) and purple (sb-UHF) diamonds.}
	\label{fig:real_UHF}
\end{figure}

\begin{figure*}
	\includegraphics[width=0.8\linewidth]{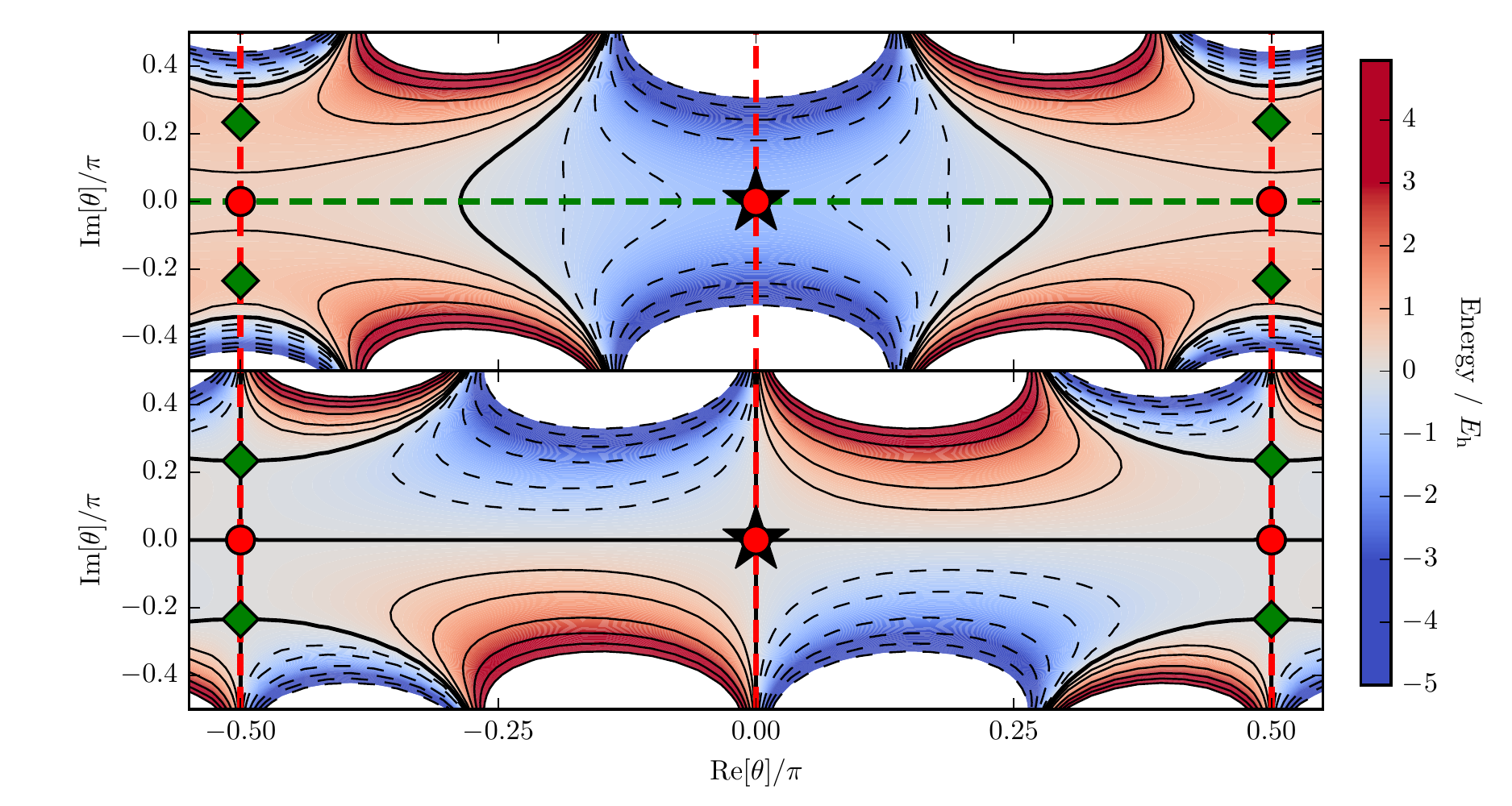}
	\caption{The real (top) and imaginary (bottom) components of the h-RHF energy of \ce{H2} at $0.75$ \r{A} bond length for $\tha = \thb = \theta$ as a function of $\Re(\theta)$ and $\Im(\theta)$.
		The parity (site-flip) operator produces the inversion symmetry $\theta \rightarrow -\theta$, as indicated by inversion through the black star.
Lines of symmetry in the vertical direction (dashed red lines) along $\Re(\theta) =0$ and $\Re(\theta) = \pm \pi/2$ coincide with the condition \eqref{eq:pt_rhf} for $\PT$-symmetry.
The energy on either side of this $\PT$-symmetry line are related by complex conjugation, while the energy along the line itself is real.
	Spatial symmetry-pure stationary points are indicated by red (RHF) circles, while complex holomorphic stationary points are illustrated by green (h-RHF) diamonds.
Note the energy is also real along the line $\Im(\theta) = 0$ (dashed green line), since this is the line of $\cK$-symmetry along which the orbital coefficients are all real.}	\label{fig:holo_RHF}
\end{figure*}

In addition to the \titou{spatially} symmetry-pure configurations $\sigg^2$, $\sigu^2$ and $\sigg\sigu$ (corresponding to two RHF and a doubly degenerate pair of UHF solutions), it is well known that, in the dissociation limit, a pair of degenerate spatial symmetry-broken UHF (sb-UHF) solutions develop (dashed purple line in Fig.~\ref{fig:H2BindingCurve}) in which the electrons localise on opposite atoms.\cite{SzaboBook}
These solutions have a form given by the parameterisation
\begin{equation}
\label{eq:UHF_param}
	\WfnUHF(\br_{1}, \br_{2}) =
	\frac{1}{\sqrt{2}}
	\begin{vmatrix}
		\phiUHF (\br_{1})\alpha(1)	&	\phiUHF (-\br_{1})\beta(1)	\\
		\phiUHF (\br_{2})\alpha(2)	&	\phiUHF (-\br_{2})\beta(2)	\\
	\end{vmatrix},
\end{equation}
where $\phiUHF$ represents the optimised spatial orbital corresponding to the UHF solution, and $\br_{1}$ and $\br_{2}$ are the spatial coordinates of electrons 1 and 2 respectively.
In the dissociation limit, the optimal UHF solutions can be represented schematically as the diradical configurations (spin-density waves)
\begin{equation}
\quad		\updotl		\quad	\dwdotr \quad	\text{and}   \quad		\dwdotl		\quad	\updotr .
\end{equation}
However, apart from the chemically intuitive idea of electron correlation, the justification for solutions existing with this particular form is not obvious.

Similarly (although less studied), a pair of degenerate spatial symmetry-broken RHF (sb-RHF) solutions develop (dashed cyan line in Fig.~\ref{fig:H2BindingCurve}) with a form given by the parameterisation
\begin{equation}
\label{eq:RHF_param}
	\WfnRHF(\br_{1}, \br_{2}) = \phiRHF(\br_{1}) \phiRHF(\br_{2}) \frac{\alpha(1) \beta(2) - \beta(1) \alpha(2)}{\sqrt{2}},
\end{equation}
where $\phiRHF$ represents the optimised spatial orbital corresponding to the RHF solution.
In the dissociation limit, the sb-RHF solution corresponds to the localisation of both electrons on the same atom to produce ionic configurations (charge-density waves)
\begin{equation}
\quad		\uddot		\quad	\vac \quad	\text{and}   \quad		\vac		\quad	\uddot .
\end{equation}
Both sb-RHF and sb-UHF solutions are extrema of the HF equations. 
However, instead of being minima like the sb-UHF solutions, the sb-RHF states correspond to maxima of the HF equations (see Fig.~\ref{fig:H2BindingCurve}).

Rather than considering the parameterisations \eqref{eq:UHF_param} and \eqref{eq:RHF_param}, we instead consider the full UHF space using two molecular orbitals 
\begin{subequations}
\begin{align}
	\phi_{\alpha}	& = \sigg \cos \tha	+ \sigu \sin \tha,
	\\
	\phi_{\beta}	& = \sigg \cos \thb	+ \sigu \sin \thb, 
\end{align}
\end{subequations}
where $\tha$ and $\thb$ are rotation angles controlling the degree of orbital mixing.
The occupied orbital coefficient matrix in the combined spatial and spinor direct product basis is therefore given by
\begin{equation}
\label{eq:UHFCoeff}
	\bC = 
	\begin{pmatrix}
		\bCa
		\\
		\bCb
	\end{pmatrix}
	=
	\begin{pmatrix}
		\cos \tha 	&	0 						
		\\
		\sin \tha 	&	0						
		\\
		0 			&	\cos \thb	
		\\
		0 			&	\sin \thb	
		\\
	\end{pmatrix}.
\end{equation}
Considering the complex-symmetric density matrix with the block form
\begin{equation}
	\bD = \bC \T{\bC} 
	= 
	\begin{pmatrix}
		\bD_{\alpha \alpha}	&	\bO
		\\
		\bO &	\bD_{\beta \beta}
	\end{pmatrix},
\end{equation}
the condition for $\PT$-symmetry [see Eq.~\eqref{eq:PTCtoD}] becomes
\begin{equation}
\label{eq:PTDenH2}
	\begin{pmatrix}
		\bD_{\alpha \alpha}	&	\bO
		\\
		\bO		&	\bD_{\beta \beta}
	\end{pmatrix}
	=
	\begin{pmatrix}
		\mP \bD_{\beta \beta}^* \mP 	&	\bO 
		\\ 
		\bO      		& \mP \bD_{\alpha \alpha}^* \mP
	\end{pmatrix}.
\end{equation}
Using the parameterisation \eqref{eq:UHFCoeff}, this condition reduces to
\begin{equation}
\label{eq:PTSymRed}
\begin{split}
	\begin{pmatrix}
		\hphantom{-\frac{1}{2}} \cos^2 \thb^* 	&	- \frac{1}{2} \sin 2 \thb^*
		\\
		- \frac{1}{2} \sin 2 \thb^*	 			&	\hphantom{-\frac{1}{2}} \sin^2 \thb^* 						
	\end{pmatrix}
	=
	\begin{pmatrix}
		\hphantom{\frac{1}{2}} \cos^2 \tha 	  	&	\frac{1}{2} \sin 2 \tha
		\\
		\frac{1}{2} \sin 2 \tha					&	\hphantom{\frac{1}{2}} \sin^2 \tha 				
	\end{pmatrix},
\end{split}
\end{equation}
which is satisfied when
\begin{equation}
	\tan(\tha) = - \tan(\thb)^* = \text{tan}(-\thb^*).
\label{eq:pt_cond2}
\end{equation}

Clearly for the case of real UHF, i.e.~$(\tha, \thb) \in \mathbb{R}$, Eq.~\eqref{eq:pt_cond2} is satisfied only when $\tha + \thb = m \pi$ for $m \in \mathbb{Z}$,
upon which we obtain the constrained molecular orbitals
\begin{subequations}
\begin{align}
	\phi_{\alpha}	& = \sigg \cos \tha +  \sigu \sin \tha,
	\\
	\phi_{\beta}	& = \sigg \cos \tha -  \sigu \sin \tha.
\end{align}
\end{subequations}
In fact, the condition for $\PT$-symmetry in real UHF aligns \textit{exactly} with the parameterisation provided by Eq.~\eqref{eq:UHF_param}, as shown in Fig.~\ref{fig:real_UHF}.
Furthermore, it is relatively simple to understand why this symmetry must exist.
Since the orbital coefficients (and by extension the density) are all real, the action of $\cT$ simply interconverts the two spin states (\textit{spin-flip}), while the spatial (parity) operator corresponds to a \textit{site-flip}.
The spin-flip already gives rise to the symmetry plane $\theta_{\alpha} = \theta_{\beta}$, along which all RHF solutions lie (cyan line in Fig.~~\ref{fig:real_UHF}).
The combined action of spin- and site-flip then essentially gives rise to the $\PT$-symmetry operation for real orbitals.

Schematically, the $\PT$ operation can be depicted as
\begin{align}
												&	\quad		\updotl		\quad	\dwdotr		
	&	\xrightarrow[\text{site-flip}]{\cP}		&	\quad		\dwdotl		\quad	\updotr		
	&	\xrightarrow[\text{spin-flip}]{\cT}		&	\quad		\updotl		\quad	\dwdotr	
	&	\cmark	
\end{align}
Therefore, in the minimal basis considered here, the (real) sb-UHF solutions (which can be labelled as diradical configurations) are $\PT$-symmetric.
In contrast, following a similar argument, the sb-RHF solutions \eqref{eq:RHF_param} (corresponding to ionic configurations) are definitely not $\PT$-symmetric:
\begin{align}
												& 	\quad		\uddot		\quad	\vac
	&	\xrightarrow[\text{site-flip}]{\cP}		&	\quad		\vac		\quad	\uddot		
	&	\xrightarrow[\text{spin-flip}]{\cT}		&	\quad		\vac		\quad	\uddot	
	&	\xmark
\end{align}
However, appropriate linear combinations of these ionic configurations (i.e.~multideterminant expansions) can be made to satisfy $\PT$-symmetry.
We note that the spatial symmetry-pure $\sigg^2$ and $\sigu^2$ states both also satisfy $\PT$-symmetry, while the $\sigg \sigu$ solutions are $\PT$-symmetry broken and are interconverted by the $\PT$ operator.

In summary, the real UHF energy surface shows $\PT$-symmetry along the line coinciding with Eq.~\eqref{eq:UHF_param}, justifying the use of this parameterisation for locating sb-UHF solutions.
As real orbital coefficients lead to purely real energies, states interconverted by this symmetry are strictly degenerate rather than being related by complex conjugation.
The well-known sb-UHF states, resembling diradical configurations, lie on this line and are $\PT$-symmetric solutions with their occupied orbitals forming a $\PT$-doublet.

\subsection{Complex Orbital Coefficients}
\label{subsec:ComplexCase}

\begin{figure*}
	\includegraphics[width=0.8\linewidth]{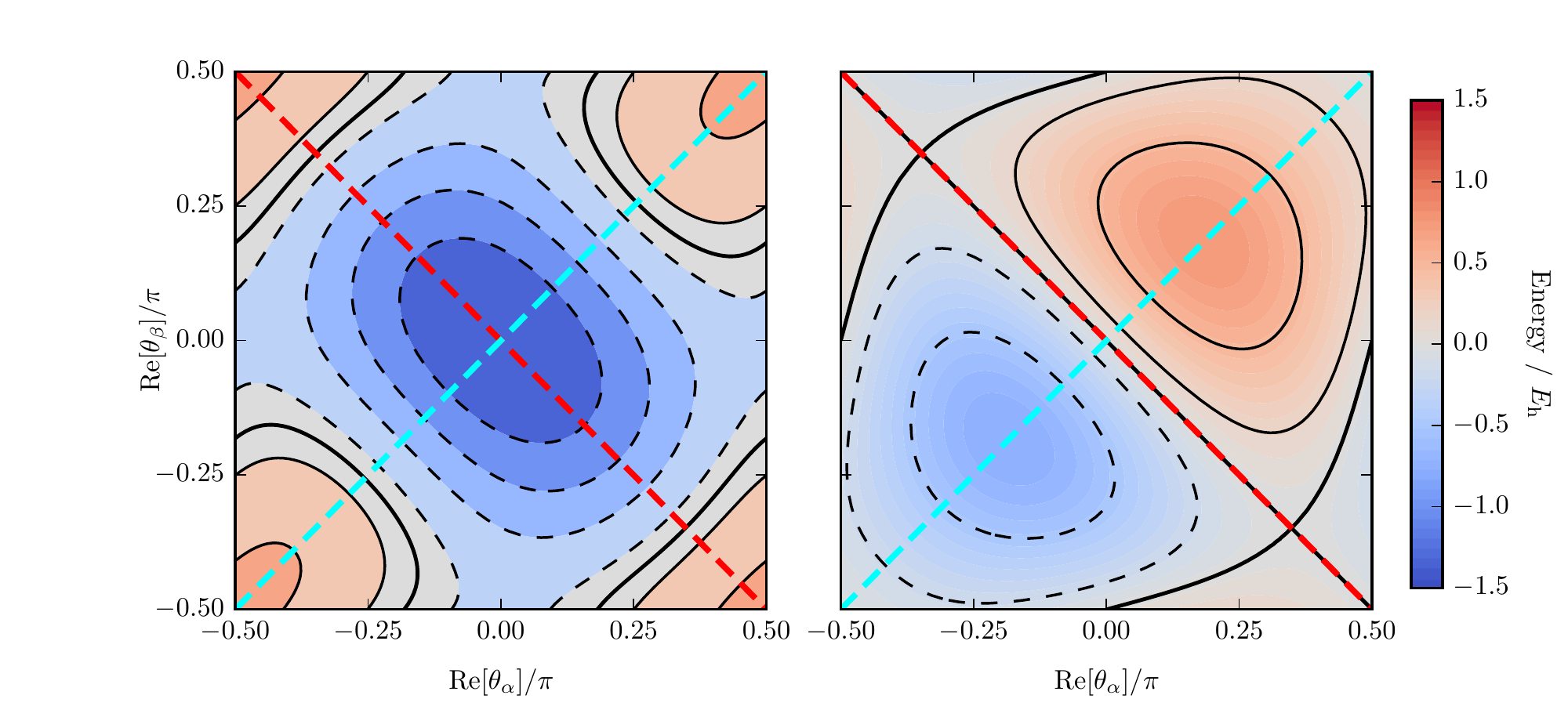}
	\caption{The real (left) and imaginary (right) components of the h-UHF energy of \ce{H2} at $0.75$ \r{A}  bond length for the illustrative case of fixed imaginary components $\Im( \tha ) = \Im( \thb ) = \pi / 8$.
Two lines of symmetry exist along $\Re(\tha) = \Re(\thb)$ and $\Re(\tha) + \Re(\thb) = \pi$, corresponding to spin-flip (dashed cyan line) and $\PT$-symmetry (dashed red line) respectively.
The energy along the $\PT$-symmetry line is real, as predicted, while the energies for states related by $\PT$-symmetry form complex-conjugate pairs.
We also note that on this illustrative landscape representing only a slice of the full holomorphic HF energy surface, there are no holomorphic HF stationary points.}
	\label{fig:holo_UHF}
\end{figure*}

Next we turn to the case of complex orbital coefficients with $(\tha, \thb) \in \mathbb{C}$.
The constraint \eqref{eq:pt_cond2} can then be decomposed into real and imaginary parts
\begin{align}
	\Re(\tha) & + \Re(\thb) = m \pi , 
	&
	\Im(\tha) & = \Im(\thb).
\end{align}
Along these lines of symmetry we expect the holomorphic HF energy to be real, while we expect the energies of density matrices interconverted by the $\PT$ operator to be related by complex conjugation.

To visualise this symmetry, we first consider the h-RHF case where $\tha = \thb =\theta$ and for which we expect $\PT$-symmetric densities when
\begin{equation}
	\theta = m \pi/2 + \I \vartheta,
\label{eq:pt_rhf}
\end{equation}
for $\vartheta \in \mathbb{R}$.
Since the holomorphic energy is complex in general, we plot the real and imaginary parts of the energy separately at a bond length of $0.75$~\r{A} as functions of $\Re(\theta)$ and $\Im(\theta)$ in Fig.~\ref{fig:holo_RHF}.
As expected, lines of $\PT$-symmetry exist along the values of $\theta$ satisfying Eq.~\eqref{eq:pt_rhf}, where the energy either side is related by complex conjugation and the energy along the line of symmetry is real.
More explicitly, the (vertical) red lines in Fig.~\ref{fig:holo_RHF} along $\Re(\theta) = 0$ and $\Re(\theta) = \pm \pi/2$ coincide with the condition \eqref{eq:pt_rhf} for $\PT$-symmetry and correspond to real energies.
The energy is also real along the line $\Im(\theta) = 0$ (green line), since this corresponds to the $\cK$-symmetry line along which the orbital coefficients are all real.
As illustrated in Fig.~\ref{fig:holo_RHF}, the h-RHF stationary solutions (green diamonds) lie on the red line, and the h-RHF states are therefore $\PT$-symmetric.
Since the h-RHF solutions are all spin-flip symmetric, we can justify $\PT$-symmetry in the energy landscape as the combination of site-flip (centre of inversion at $\theta = 0$) and complex conjugation.

Finally, we consider the complex h-UHF case. 
As the h-UHF energy is a function of four real variables (real and imaginary parts of $\tha$ and $\thb$), we illustrate the general symmetry using the specific case $\Im(\tha) = \Im(\thb) = \pi / 8$ and visualise the energy for \ce{H2} at a bond length of $0.75$~\r{A} in Fig.~\ref{fig:holo_UHF}.
We find the lines of spin-flip (cyan) and $\PT$-symmetry (red) occur as in Fig.~\ref{fig:real_UHF}, where now the complex-conjugation of energies on either side of the line of $\PT$-symmetry is explicitly observable.
Since the site-flip operation takes $(\tha, \thb) \rightarrow (-\tha, -\thb)$, we cannot observe its effect on the energy landscape under the constraint $\Im(\tha) = \Im(\thb) = \pi / 8$, and we note that there are no h-UHF stationary points in Fig.~\ref{fig:real_UHF}.
Moreover, inspection of the stationary points corresponding to the h-UHF solutions in Fig.~\ref{fig:H2BindingCurve} reveals that they follow the form $(\tha, \thb) = ( \I \vartheta, - \I \vartheta)$, for $\vartheta \in \mathbb{R}$, leading to the conclusion that the h-UHF solutions in \ce{H2} \textit{are not} $\PT$-symmetric.

In summary, the $\PT$-symmetric stationary h-HF solutions for \ce{H2} are the $\sigg^2$, $\sigu^2$, sb-UHF and h-RHF states, although the effect of $\PT$-symmetry can be observed throughout the holomorphic HF energy landscape.
For these $\PT$-symmetric solutions, the molecular orbital coefficients possess the $\PT$-doublet form [see Eq.~\eqref{eq:PTDoubletC}].
Stationary points that do not correspond to $\PT$-symmetric states (including the sb-RHF and h-UHF solutions) occur in pairs which are interconverted by the action of $\PT$, and the onset of $\PT$-symmetry breaking coincides with the disappearance of the h-RHF or the sb-UHF solutions at Coulson--Fischer (quasi-exceptional) points \cite{Burton_2019} in a similar manner to other types of symmetry-breaking in HF theory. \cite{Fukutome_1981, StuberPaldus, Jimenez-Hoyos_2011}

\section{Concluding remarks}

In this work, we have outlined the conditions for $\PT$-symmetry --- a weaker condition than Hermiticity which ensures real energies --- in electronic structure, and specifically \titou{the HF approximation}.
\titou{In particular, we have explored the existence of $\PT$-symmetry in the non-Hermitian h-HF formulation that forms the rigorous analytic continuation of real HF.}
Our most important results are:
\begin{enumerate}

	\item A set of molecular orbitals is $\PT$-symmetric \textit{if and only if} the effective Fock Hamiltonian is $\PT$-symmetric, and \textit{vice versa}.

	\item Starting with a $\PT$-symmetric guess density matrix, $\PT$-symmetry can be conserved throughout the self-consistent process.

	\item If an optimal self-consistent solution is invariant under $\PT$, then its eigenvalues and corresponding HF energy must be real.

	\item $\PT$-symmetry can be explicitly satisfied by constructing the molecular orbitals coefficients in the structure of a so-called $\PT$-doublet, i.e.~pairing each occupied orbital with its $\PT$-transformed analogue.

	\item Slater determinants built from $\PT$-doublets lead to $\PT$-symmetric many-electron wave functions.
\end{enumerate}

$\PT$-symmetry provides a novel intrinsic symmetry in the HF energy landscape, where the energies of densities interconverted by $\PT$ are related by complex conjugation.
For real HF, this symmetry corresponds to the combination of the parity and spin-flip operations.
As an illustrative example, we have considered the \ce{H2} molecule in a minimal basis, where we have observed the effects of $\PT$-symmetry on the HF energy landscape. 
In particular, we have found that the sb-UHF and h-RHF wave functions are $\PT$-symmetric, while the sb-RHF and h-UHF wave functions break $\PT$-symmetry but occur in complex conjugate pairs related by the $\PT$ operator.
The transitions between broken and unbroken $\PT$-symmetry regions coincide with the disappearance of the h-RHF or the sb-UHF solutions at Coulson--Fischer points.

By demonstrating the existence of $\PT$-symmetric solutions with real energies in the HF approximation, we remove the rigorous condition of Hermiticity that is usually applied in electronic structure theory.
We are currently working on the implementation of restricted, unrestricted and generalised HF self-consistent approaches that explicitly enforce this symmetry.
Ultimately, by bridging the gap between $\PT$-symmetric physics and quantum chemistry, we hope to pave the way for the development of new classes of non-Hermitian Hamiltonians with real eigenvalues in electronic structure theory.

\begin{acknowledgements}
H.G.A.B.~thanks the Cambridge Trust for a studentship and A.J.W.T.~thanks the Royal Society for a University Research Fellowship (UF110161). 
We also thank Bang~Huynh for insightful conversations throughout the development of this work.
\end{acknowledgements}

\appendix
\section{\titou{$\cT$-Symmetry for General Spins}}
\label{A:GeneralSpin}
We loosely follow Weinberg's discussion on the nature of $\cT$ in Ref.~\onlinecite{WeinbergBook}, although we use notation more familiar to electronic structure.
Since the action of $\cT$ reverses angular momentum, we require spin-angular momentum operators $\bopS$ to satisfy
\begin{align}
\bopS \cT  &= - \cT \bopS, &  \bopS^2 \cT  &=  \cT \bopS^2.
\label{eq:S}
\end{align}
Considering a general spin state $\ket{\s, \ms}$, we can then show
\begin{align}
\opSz \qty[ \cT \ket{\s, \ms} ] &= - \cT \opSz \ket{\s, \ms} = - \ms \qty[ \cT \ket{\s, \ms} ] ,
\\
\bopS^2 \qty[  \cT \ket{\s, \ms} ]  &= \cT \bopS^2 \ket{\s, \ms} = \s (\s +1) \qty[ \cT \ket{\s, \ms} ]. 
\end{align}
In combination, these results imply
\begin{equation}
\cT \ket{\s, \ms } = \cphase(\s, \ms) \ket{\s, -\ms},
\label{eq:cTaction}
\end{equation}
for some complex value $\cphase(\s, \ms)$.

To identify the functional form of $\cphase(\s, \ms)$, we use the ladder operators $\opS_{\pm} = \opSx \pm \I \opSy$ which, due to the anti-linear character of $\cT$, satisfy
\begin{equation}
\opSpm \cT = - \cT \opSmp.
\end{equation}
From the standard ladder operator relationship
\begin{equation}
\opSpm = \ladphase_{\pm}(\s, \ms) \ket{\s, \ms \pm 1}, 
\end{equation}
where $\ladphase_{\pm}(\s, \ms) = \sqrt{\s (\s +1 ) - \ms (\ms \pm 1)}$,
we find
\begin{align}
\begin{split}
\opSpm \cT \ket{\s, \ms} 
&= - \cT \opSmp \ket{\s, \ms} 
\\
&= - \ladphase_{\mp}(\s, \ms)  \cT  \ket{\s, \ms \mp 1}.
\end{split}
\label{eq:LadOpResult}
\end{align}
Alternatively, from the result of Eq.~\eqref{eq:cTaction} we can explicitly identify 
\begin{equation}
\opSpm \cT \ket{\s, \ms} = \cphase(\s, \ms)  \ladphase_{\pm}(\s, -\ms) \ket{\s, -\ms \pm 1}, 
\label{eq:LHS}
\end{equation}
and
\begin{equation}
\cT  \ket{\s, \ms \mp 1} = \cphase(\s, \ms \mp 1) \ket{\s, -\ms \pm 1}.
\label{eq:RHS}
\end{equation}
Inserting Eq.~\eqref{eq:LHS} and Eq.~\eqref{eq:RHS} into the LHS and RHS of Eq.~\eqref{eq:LadOpResult} respectively, and noting that $\ladphase_{\pm}(\s, -\ms) = \ladphase_{\mp}(\s, \ms)$, we find
\begin{equation}
\cphase(\s, \ms) = -  \cphase(\s, \ms \mp 1) = (-1)^{\s - \ms} \cphase(\s, \s).
\end{equation}
Here $\cphase(\s, \s)$ is an arbitrary complex value that is conventionally set to unity.
As a result, the action of $\cT$ on a function $\phi$, expressed in the basis $\qty{ \ket{\s, \ms} }$ with dimension $\ns = (2\s + 1)$, is given by
\begin{equation}
\cT \phi = \bZ \cK \phi,
\end{equation}
where $\bZ$ is an $(\ns \times \ns)$ orthogonal matrix representing the action of $\cT$ on the spin eigenfunctions, given explicitly as
\begin{equation}
Z_{ij} = 
\begin{cases}
(-1)^{2\s+1 - i},	& \text{if } i + j = 2 (\s+1), 
\\
0,  				& \text{otherwise.}
\end{cases}
\end{equation} 

Next, consider the specific $\bZ$ matrices for various spin cases:
\begin{align*}
\s = 0 &: \bZ = \begin{pmatrix} 1 \end{pmatrix},
\\
\s = \frac{1}{2} &: \bZ = 
\begin{pmatrix}
0 & 1 
\\
-1 & 0
\end{pmatrix},
\\
\s = 1 &: \bZ = 
\begin{pmatrix}
0 & 0 & 1 
\\
0 & -1 & 0
\\
1 & 0 & 0
\end{pmatrix},
\\
\s = \frac{3}{2} &: \bZ = 
\begin{pmatrix}
0 & 0 & 0 & 1
\\
0 & 0 & -1 & 0
\\
0 & 1 & 0 & 0
\\
-1 & 0 & 0 & 0
\end{pmatrix},
\\
\s = 2 &: \bZ = 
\begin{pmatrix}
0 & 0 & 0 & 0 & 1 
\\
0 & 0 & 0 & -1 & 0 
\\ 
0 & 0 & 1 & 0 & 0 
\\ 
0 & -1 & 0 & 0 & 0 
\\
1 & 0 & 0 & 0 & 0
\\
\end{pmatrix}.
\\
&\vdots 
\end{align*}
Significantly, for the bosonic (integer $\s$) case, $\bZ$ is symmetric, i.e. $\bZ = \T{\bZ}$, while in the fermionic  (half-integer $\s$) case, $\bZ$ becomes skew-symmetric, i.e. $\bZ = -\T{\bZ}$. 
This observation leads to two key results.
First, considering the operation $\cT^{2} \phi = \bZ^2 \cK^2 \phi$, we find $\bZ^2 = \T{\bZ} \bZ = \bI$ in the bosonic case and $\bZ^2 = -\T{\bZ} \bZ = - \bI$ for the fermionic case.
When combined with the relationship $\cK^2 = \cI$, this result leads directly to the even and odd character of $\cT$ for bosons and fermions respectively.
Secondly, since $\bZ$ is symmetric and orthogonal in the bosonic case, it can be decomposed into the form $\bZ = \bV \bSig \T{\bV}$, where $\bV$ is orthogonal and $\bSig$ is a diagonal matrix containing the eigenvalues of $\bZ$, each equal to $-1$ or $+1$.
Taking $\bSig = \blam \T{\blam}$, we find
\begin{equation}
\cT \phi = \bZ \cK \phi = \bV \blam \T{\blam} \T{\bV} \cK \phi = (\bV \blam ) \cK (\bV \blam )^{\dag} \phi.
\end{equation} 
Consequently, for the bosonic case, one can always find a transformation $(\bV \blam)$ into a basis under which the action of time-reversal reduces to $\cT = \cK$, described in Ref.~\onlinecite{Jones-Smith_2010} as a ``canonical'' bosonic basis.
Similarly, in the fermionic case, the properties of skew-symmetric orthogonal matrices allow $\bZ$ to be decomposed into the form $\bZ = \bQ \bLam \T{\bQ}$,\cite{Zumino1962} where $\bQ$ is orthogonal and $\bLam$ takes the form
\begin{equation}
\bLam = 
\begin{pmatrix}
\I \sigy 	& 		  	& 
\\
         	& \ddots 	&
\\
			&			& \I \sigy
\end{pmatrix}.
\end{equation}
As a result, the action of $\cT$ can be expressed as  
\begin{equation}
\cT \phi = \bZ \cK \phi = \bQ \bLam \T{\bQ} \cK \phi = (\bQ) \bLam \cK (\bQ)^{\dag} \phi,
\end{equation} 
and thus, for fermionic systems, it is always possible to find a transformation $\bQ$ into a canonical fermionic basis in which the action of time-reversal reduces to $\cT = \bLam \cK$.\cite{Jones-Smith_2010} 

\bibliography{ptHF,ptHF-control}

\providecommand{\latin}[1]{#1}
\makeatletter
\providecommand{\doi}
  {\begingroup\let\do\@makeother\dospecials
  \catcode`\{=1 \catcode`\}=2 \doi@aux}
\providecommand{\doi@aux}[1]{\endgroup\texttt{#1}}
\makeatother
\providecommand*\mcitethebibliography{\thebibliography}
\csname @ifundefined\endcsname{endmcitethebibliography}
  {\let\endmcitethebibliography\endthebibliography}{}
\begin{mcitethebibliography}{79}
\providecommand*\natexlab[1]{#1}
\providecommand*\mciteSetBstSublistMode[1]{}
\providecommand*\mciteSetBstMaxWidthForm[2]{}
\providecommand*\mciteBstWouldAddEndPuncttrue
  {\def\EndOfBibitem{\unskip.}}
\providecommand*\mciteBstWouldAddEndPunctfalse
  {\let\EndOfBibitem\relax}
\providecommand*\mciteSetBstMidEndSepPunct[3]{}
\providecommand*\mciteSetBstSublistLabelBeginEnd[3]{}
\providecommand*\EndOfBibitem{}
\mciteSetBstSublistMode{f}
\mciteSetBstMaxWidthForm{subitem}{(\alph{mcitesubitemcount})}
\mciteSetBstSublistLabelBeginEnd
  {\mcitemaxwidthsubitemform\space}
  {\relax}
  {\relax}

\bibitem[Szabo and Ostlund(1989)Szabo, and Ostlund]{SzaboBook}
Szabo,~A.; Ostlund,~N.~S. \emph{Modern quantum chemistry}; McGraw-Hill: New
  York, 1989\relax
\mciteBstWouldAddEndPuncttrue
\mciteSetBstMidEndSepPunct{\mcitedefaultmidpunct}
{\mcitedefaultendpunct}{\mcitedefaultseppunct}\relax
\EndOfBibitem
\bibitem[Parr and Yang(1989)Parr, and Yang]{ParrBook}
Parr,~R.~G.; Yang,~W. \emph{Density-functional theory of atoms and molecules};
  Oxford: Clarendon Press, 1989\relax
\mciteBstWouldAddEndPuncttrue
\mciteSetBstMidEndSepPunct{\mcitedefaultmidpunct}
{\mcitedefaultendpunct}{\mcitedefaultseppunct}\relax
\EndOfBibitem
\bibitem[Coulson and Fischer(1949)Coulson, and Fischer]{Coulson_1949}
Coulson,~C.; Fischer,~I. {XXXIV. Notes on the Molecular Orbital Treatment of
  the Hydrogen Molecule}. \emph{Philos. Mag.} \textbf{1949}, \emph{40},
  386\relax
\mciteBstWouldAddEndPuncttrue
\mciteSetBstMidEndSepPunct{\mcitedefaultmidpunct}
{\mcitedefaultendpunct}{\mcitedefaultseppunct}\relax
\EndOfBibitem
\bibitem[Giuliani and Vignale(2005)Giuliani, and Vignale]{VignaleBook}
Giuliani,~G.~F.; Vignale,~G. \emph{Quantum theory of the electron liquid};
  Cambridge University Press: Cambridge, 2005\relax
\mciteBstWouldAddEndPuncttrue
\mciteSetBstMidEndSepPunct{\mcitedefaultmidpunct}
{\mcitedefaultendpunct}{\mcitedefaultseppunct}\relax
\EndOfBibitem
\bibitem[Jimenez-Hoyos \latin{et~al.}(2012)Jimenez-Hoyos, Henderson,
  Tsuchimochi, and Scuseria]{Jimenez-Hoyos_2012}
Jimenez-Hoyos,~C.~A.; Henderson,~T.~M.; Tsuchimochi,~T.; Scuseria,~G.~E.
  Projected Hartree-Fock Theory. \emph{J. Chem. Phys.} \textbf{2012},
  \emph{136}, 164109\relax
\mciteBstWouldAddEndPuncttrue
\mciteSetBstMidEndSepPunct{\mcitedefaultmidpunct}
{\mcitedefaultendpunct}{\mcitedefaultseppunct}\relax
\EndOfBibitem
\bibitem[Cui \latin{et~al.}(2013)Cui, Bulik, Jimenez-Hoyos, Henderson, and
  Scuseria]{Cui_2013}
Cui,~Y.; Bulik,~I.~W.; Jimenez-Hoyos,~C.~A.; Henderson,~T.~M.; Scuseria,~G.~E.
  Proper and improper zero energy modes in Hartree-Fock theory and their
  relevance for symmetry breaking and restoration. \emph{J. Chem. Phys.}
  \textbf{2013}, \emph{139}, 154107\relax
\mciteBstWouldAddEndPuncttrue
\mciteSetBstMidEndSepPunct{\mcitedefaultmidpunct}
{\mcitedefaultendpunct}{\mcitedefaultseppunct}\relax
\EndOfBibitem
\bibitem[Qiu \latin{et~al.}(2017)Qiu, Henderson, Zhao, and Scuseria]{Qiu_2017}
Qiu,~Y.; Henderson,~T.~M.; Zhao,~J.; Scuseria,~G.~E. Projected coupled cluster
  theory. \emph{J. Chem. Phys.} \textbf{2017}, \emph{147}, 064111\relax
\mciteBstWouldAddEndPuncttrue
\mciteSetBstMidEndSepPunct{\mcitedefaultmidpunct}
{\mcitedefaultendpunct}{\mcitedefaultseppunct}\relax
\EndOfBibitem
\bibitem[Jake \latin{et~al.}(2018)Jake, Henderson, and Scuseria]{Jake_2018}
Jake,~L.~C.; Henderson,~T.~M.; Scuseria,~G.~E. Hartree\textendash{{Fock}}
  Symmetry Breaking around Conical Intersections. \emph{J. Chem. Phys.}
  \textbf{2018}, \emph{148}, 024109\relax
\mciteBstWouldAddEndPuncttrue
\mciteSetBstMidEndSepPunct{\mcitedefaultmidpunct}
{\mcitedefaultendpunct}{\mcitedefaultseppunct}\relax
\EndOfBibitem
\bibitem[Lykos and Pratt(1963)Lykos, and Pratt]{Lykos_1963}
Lykos,~P.; Pratt,~G.~W. Discussion on {{The Hartree}}-{{Fock Approximation}}.
  \emph{Rev. Mod. Phys.} \textbf{1963}, \emph{35}, 496--501\relax
\mciteBstWouldAddEndPuncttrue
\mciteSetBstMidEndSepPunct{\mcitedefaultmidpunct}
{\mcitedefaultendpunct}{\mcitedefaultseppunct}\relax
\EndOfBibitem
\bibitem[Fukutome()]{Fukutome_1981}
Fukutome,~H. Unrestricted {Hartree--Fock} theory and its applications to
  molecules and chemical reactions. \emph{Int. J. Quantum Chem.} 955\relax
\mciteBstWouldAddEndPuncttrue
\mciteSetBstMidEndSepPunct{\mcitedefaultmidpunct}
{\mcitedefaultendpunct}{\mcitedefaultseppunct}\relax
\EndOfBibitem
\bibitem[Stuber and Paldus(2003)Stuber, and Paldus]{StuberPaldus}
Stuber,~J.; Paldus,~J. {Symmetry Breaking in the Independent Particle Model}.
  In \emph{Fundamental World of Quantum Chemistry: A Tribute to the Memory of
  Per-Olov L\"{o}wdin}; Br\"{a}ndas,~E.~J., Kryachko,~E.~S., Eds.; Kluwer
  Academic: Dordrecht, 2003; Vol.~1; p~67\relax
\mciteBstWouldAddEndPuncttrue
\mciteSetBstMidEndSepPunct{\mcitedefaultmidpunct}
{\mcitedefaultendpunct}{\mcitedefaultseppunct}\relax
\EndOfBibitem
\bibitem[{Jim\'{e}nez-Hoyos} \latin{et~al.}(2011){Jim\'{e}nez-Hoyos},
  Henderson, and Scuseria]{Jimenez-Hoyos_2011}
{Jim\'{e}nez-Hoyos},~C.~A.; Henderson,~T.~M.; Scuseria,~G.~E. Generalized
  Hartree--Fock Description of Molecular Dissociation. \emph{J. Chem. Theory
  Comput.} \textbf{2011}, \emph{7}, 2667\relax
\mciteBstWouldAddEndPuncttrue
\mciteSetBstMidEndSepPunct{\mcitedefaultmidpunct}
{\mcitedefaultendpunct}{\mcitedefaultseppunct}\relax
\EndOfBibitem
\bibitem[Hiscock and Thom(2014)Hiscock, and Thom]{Hiscock_2014}
Hiscock,~H.~G.; Thom,~A. J.~W. Holomorphic {{Hartree}}\textendash{{Fock
  Theory}} and {{Configuration Interaction}}. \emph{J. Chem. Theory Comput.}
  \textbf{2014}, \emph{10}, 4795--4800\relax
\mciteBstWouldAddEndPuncttrue
\mciteSetBstMidEndSepPunct{\mcitedefaultmidpunct}
{\mcitedefaultendpunct}{\mcitedefaultseppunct}\relax
\EndOfBibitem
\bibitem[Burton and Thom(2016)Burton, and Thom]{Burton_2016}
Burton,~H. G.~A.; Thom,~A. J.~W. Holomorphic {Hartree--Fock} Theory: An
  Inherently Multireference Approach. \emph{J. Chem. Theory Comput.}
  \textbf{2016}, \emph{12}, 167\relax
\mciteBstWouldAddEndPuncttrue
\mciteSetBstMidEndSepPunct{\mcitedefaultmidpunct}
{\mcitedefaultendpunct}{\mcitedefaultseppunct}\relax
\EndOfBibitem
\bibitem[Burton \latin{et~al.}(2018)Burton, Gross, and Thom]{Burton_2018}
Burton,~H. G.~A.; Gross,~M.; Thom,~A. J.~W. Holomorphic
  {{Hartree}}\textendash{{Fock Theory}}: {{The Nature}} of {{Two}}-{{Electron
  Problems}}. \emph{J. Chem. Theory Comput.} \textbf{2018}, \emph{14},
  607--618\relax
\mciteBstWouldAddEndPuncttrue
\mciteSetBstMidEndSepPunct{\mcitedefaultmidpunct}
{\mcitedefaultendpunct}{\mcitedefaultseppunct}\relax
\EndOfBibitem
\bibitem[Moiseyev(2011)]{MoiseyevBook}
Moiseyev,~N. \emph{Non-Hermitian Quantum Mechanics}; Cambridge University
  Press, 2011\relax
\mciteBstWouldAddEndPuncttrue
\mciteSetBstMidEndSepPunct{\mcitedefaultmidpunct}
{\mcitedefaultendpunct}{\mcitedefaultseppunct}\relax
\EndOfBibitem
\bibitem[Burton \latin{et~al.}(2019)Burton, Thom, and Loos]{Burton_2019}
Burton,~H. G.~A.; Thom,~A. J.~W.; Loos,~P.~F. Complex Adiabatic Connection: a
  Hidden Non-Hermitian Path from Ground to Excited States. \emph{J. Chem.
  Phys.} \textbf{2019}, \emph{150}, 041103\relax
\mciteBstWouldAddEndPuncttrue
\mciteSetBstMidEndSepPunct{\mcitedefaultmidpunct}
{\mcitedefaultendpunct}{\mcitedefaultseppunct}\relax
\EndOfBibitem
\bibitem[Seidl \latin{et~al.}(2018)Seidl, Giarrusso, Vuckovic, Fabiano, and
  Gori-Giorgi]{Seidl_2018}
Seidl,~M.; Giarrusso,~S.; Vuckovic,~S.; Fabiano,~E.; Gori-Giorgi,~P.
  Communication: Strong-interaction limit of an adiabatic connection in
  {Hartree--Fock} theory. \emph{J. Chem. Phys.} \textbf{2018}, \emph{149},
  241101\relax
\mciteBstWouldAddEndPuncttrue
\mciteSetBstMidEndSepPunct{\mcitedefaultmidpunct}
{\mcitedefaultendpunct}{\mcitedefaultseppunct}\relax
\EndOfBibitem
\bibitem[Seidl(2007)]{Seidl_2007}
Seidl,~M. Adiabatic Connection in Density-Functional Theory: {{Two}} Electrons
  on the Surface of a Sphere. \emph{Phys. Rev. A} \textbf{2007}, \emph{75},
  062506\relax
\mciteBstWouldAddEndPuncttrue
\mciteSetBstMidEndSepPunct{\mcitedefaultmidpunct}
{\mcitedefaultendpunct}{\mcitedefaultseppunct}\relax
\EndOfBibitem
\bibitem[Loos and Gill(2009)Loos, and Gill]{Loos_2009a}
Loos,~P.~F.; Gill,~P. M.~W. Ground state of two electrons on a sphere.
  \emph{Phys. Rev. A} \textbf{2009}, \emph{79}, 062517\relax
\mciteBstWouldAddEndPuncttrue
\mciteSetBstMidEndSepPunct{\mcitedefaultmidpunct}
{\mcitedefaultendpunct}{\mcitedefaultseppunct}\relax
\EndOfBibitem
\bibitem[Loos and Gill(2009)Loos, and Gill]{Loos_2009c}
Loos,~P.~F.; Gill,~P. M.~W. Two Electrons on a Hypersphere: A Quasiexactly
  Solvable Model. \emph{Phys. Rev. Lett.} \textbf{2009}, \emph{103},
  123008\relax
\mciteBstWouldAddEndPuncttrue
\mciteSetBstMidEndSepPunct{\mcitedefaultmidpunct}
{\mcitedefaultendpunct}{\mcitedefaultseppunct}\relax
\EndOfBibitem
\bibitem[Loos and Bressanini(2015)Loos, and Bressanini]{Loos_2015}
Loos,~P.-F.; Bressanini,~D. Nodal Surfaces and Interdimensional Degeneracies.
  \emph{J. Chem. Phys.} \textbf{2015}, \emph{142}, 214112\relax
\mciteBstWouldAddEndPuncttrue
\mciteSetBstMidEndSepPunct{\mcitedefaultmidpunct}
{\mcitedefaultendpunct}{\mcitedefaultseppunct}\relax
\EndOfBibitem
\bibitem[Loos \latin{et~al.}(2018)Loos, Romaniello, and Berger]{Loos_2018a}
Loos,~P.~F.; Romaniello,~P.; Berger,~J.~A. Green Functions and
  Self-Consistency: Insights From the Spherium Model. \emph{J. Chem. Theory
  Comput.} \textbf{2018}, \emph{14}, 3071\relax
\mciteBstWouldAddEndPuncttrue
\mciteSetBstMidEndSepPunct{\mcitedefaultmidpunct}
{\mcitedefaultendpunct}{\mcitedefaultseppunct}\relax
\EndOfBibitem
\bibitem[Bender and Boettcher(1998)Bender, and Boettcher]{Bender_1998}
Bender,~C.~M.; Boettcher,~S. Real Spectra in {non-Hermitian Hamiltonians}
  Having PT Symmetry. \emph{Phys. Rev. Lett.} \textbf{1998}, \emph{80},
  5243\relax
\mciteBstWouldAddEndPuncttrue
\mciteSetBstMidEndSepPunct{\mcitedefaultmidpunct}
{\mcitedefaultendpunct}{\mcitedefaultseppunct}\relax
\EndOfBibitem
\bibitem[Bender \latin{et~al.}(1999)Bender, Boettcher, and
  Meisinger]{Bender_1999}
Bender,~C.~M.; Boettcher,~S.; Meisinger,~P.~N. {{PT}}-Symmetric Quantum
  Mechanics. \emph{J. Math. Phys.} \textbf{1999}, \emph{40}, 2201--2229\relax
\mciteBstWouldAddEndPuncttrue
\mciteSetBstMidEndSepPunct{\mcitedefaultmidpunct}
{\mcitedefaultendpunct}{\mcitedefaultseppunct}\relax
\EndOfBibitem
\bibitem[Bender \latin{et~al.}(2002)Bender, Berry, and Mandilara]{Bender_2002}
Bender,~C.~M.; Berry,~M.~V.; Mandilara,~A. Generalized {{PT}} Symmetry and Real
  Spectra. \emph{J. Phys. Math. Gen.} \textbf{2002}, \emph{35},
  L467--L471\relax
\mciteBstWouldAddEndPuncttrue
\mciteSetBstMidEndSepPunct{\mcitedefaultmidpunct}
{\mcitedefaultendpunct}{\mcitedefaultseppunct}\relax
\EndOfBibitem
\bibitem[Bender \latin{et~al.}(2002)Bender, Brody, and Jones]{Bender_2002a}
Bender,~C.~M.; Brody,~D.~C.; Jones,~H.~F. Complex {{Extension}} of {{Quantum
  Mechanics}}. \emph{Phys. Rev. Lett.} \textbf{2002}, \emph{89}, 270401\relax
\mciteBstWouldAddEndPuncttrue
\mciteSetBstMidEndSepPunct{\mcitedefaultmidpunct}
{\mcitedefaultendpunct}{\mcitedefaultseppunct}\relax
\EndOfBibitem
\bibitem[Bender \latin{et~al.}(2003)Bender, Brody, and Jones]{Bender_2003}
Bender,~C.~M.; Brody,~D.~C.; Jones,~H.~F. Must a {{Hamiltonian}} Be
  {{Hermitian}}? \emph{Am. J. Phys.} \textbf{2003}, \emph{71}, 1095--1102\relax
\mciteBstWouldAddEndPuncttrue
\mciteSetBstMidEndSepPunct{\mcitedefaultmidpunct}
{\mcitedefaultendpunct}{\mcitedefaultseppunct}\relax
\EndOfBibitem
\bibitem[Bender \latin{et~al.}(2004)Bender, Brod, Refig, and
  Reuter]{Bender_2004}
Bender,~C.~M.; Brod,~J.; Refig,~A.; Reuter,~M.~E. The {{\cal C}} Operator in
  {{\cal PT}}-Symmetric Quantum Theories. \emph{J. Phys. Math. Gen.}
  \textbf{2004}, \emph{37}, 10139--10165\relax
\mciteBstWouldAddEndPuncttrue
\mciteSetBstMidEndSepPunct{\mcitedefaultmidpunct}
{\mcitedefaultendpunct}{\mcitedefaultseppunct}\relax
\EndOfBibitem
\bibitem[Bender(2005)]{Bender_2005}
Bender,~C.~M. Introduction to {{\cal PT}}-Symmetric Quantum Theory.
  \emph{Contemp. Phys.} \textbf{2005}, \emph{46}, 277--292\relax
\mciteBstWouldAddEndPuncttrue
\mciteSetBstMidEndSepPunct{\mcitedefaultmidpunct}
{\mcitedefaultendpunct}{\mcitedefaultseppunct}\relax
\EndOfBibitem
\bibitem[Bender \latin{et~al.}(2006)Bender, Chen, and Milton]{Bender_2006}
Bender,~C.~M.; Chen,~J.-H.; Milton,~K.~A. {{\cal PT}}-Symmetric versus
  {{Hermitian}} Formulations of Quantum Mechanics. \emph{J. Phys. Math. Gen.}
  \textbf{2006}, \emph{39}, 1657--1668\relax
\mciteBstWouldAddEndPuncttrue
\mciteSetBstMidEndSepPunct{\mcitedefaultmidpunct}
{\mcitedefaultendpunct}{\mcitedefaultseppunct}\relax
\EndOfBibitem
\bibitem[Bender(2007)]{Bender_2007}
Bender,~C.~M. Making Sense of Non-{{Hermitian Hamiltonians}}. \emph{Rep. Prog.
  Phys.} \textbf{2007}, \emph{70}, 947--1018\relax
\mciteBstWouldAddEndPuncttrue
\mciteSetBstMidEndSepPunct{\mcitedefaultmidpunct}
{\mcitedefaultendpunct}{\mcitedefaultseppunct}\relax
\EndOfBibitem
\bibitem[Bender \latin{et~al.}(2007)Bender, Brody, Jones, and
  Meister]{Bender_2007a}
Bender,~C.~M.; Brody,~D.~C.; Jones,~H.~F.; Meister,~B.~K. Faster than
  {{Hermitian Quantum Mechanics}}. \emph{Phys. Rev. Lett.} \textbf{2007},
  \emph{98}, 040403\relax
\mciteBstWouldAddEndPuncttrue
\mciteSetBstMidEndSepPunct{\mcitedefaultmidpunct}
{\mcitedefaultendpunct}{\mcitedefaultseppunct}\relax
\EndOfBibitem
\bibitem[Bender \latin{et~al.}(2008)Bender, Brody, and Hook]{Bender_2008}
Bender,~C.~M.; Brody,~D.~C.; Hook,~D.~W. Quantum Effects in Classical Systems
  Having Complex Energy. \emph{J. Phys. Math. Theor.} \textbf{2008}, \emph{41},
  352003\relax
\mciteBstWouldAddEndPuncttrue
\mciteSetBstMidEndSepPunct{\mcitedefaultmidpunct}
{\mcitedefaultendpunct}{\mcitedefaultseppunct}\relax
\EndOfBibitem
\bibitem[Bender and Jones(2008)Bender, and Jones]{Bender_2008a}
Bender,~C.~M.; Jones,~H.~F. Interactions of {{Hermitian}} and Non-{{Hermitian
  Hamiltonians}}. \emph{J. Phys. Math. Theor.} \textbf{2008}, \emph{41},
  244006\relax
\mciteBstWouldAddEndPuncttrue
\mciteSetBstMidEndSepPunct{\mcitedefaultmidpunct}
{\mcitedefaultendpunct}{\mcitedefaultseppunct}\relax
\EndOfBibitem
\bibitem[Bender and Jones(2014)Bender, and Jones]{Bender_2014}
Bender,~C.~M.; Jones,~H.~F. Calculation of Low-Lying Energy Levels in Quantum
  Mechanics. \emph{J. Phys. Math. Theor.} \textbf{2014}, \emph{47},
  395303\relax
\mciteBstWouldAddEndPuncttrue
\mciteSetBstMidEndSepPunct{\mcitedefaultmidpunct}
{\mcitedefaultendpunct}{\mcitedefaultseppunct}\relax
\EndOfBibitem
\bibitem[Bender(2015)]{Bender_2015}
Bender,~C.~M. {{{\cal PT}}}-Symmetric Quantum Theory. \emph{J. Phys. Conf.
  Ser.} \textbf{2015}, \emph{631}, 012002\relax
\mciteBstWouldAddEndPuncttrue
\mciteSetBstMidEndSepPunct{\mcitedefaultmidpunct}
{\mcitedefaultendpunct}{\mcitedefaultseppunct}\relax
\EndOfBibitem
\bibitem[Bender(2016)]{Bender_2016}
Bender,~C.~M. {{PT}} Symmetry in Quantum Physics: {{From}} a Mathematical
  Curiosity to Optical Experiments. \emph{Europhys. News} \textbf{2016},
  \emph{47}, 17--20\relax
\mciteBstWouldAddEndPuncttrue
\mciteSetBstMidEndSepPunct{\mcitedefaultmidpunct}
{\mcitedefaultendpunct}{\mcitedefaultseppunct}\relax
\EndOfBibitem
\bibitem[Bender \latin{et~al.}(2017)Bender, Hassanpour, Hook, Klevansky,
  S\"underhauf, and Wen]{Bender_2017}
Bender,~C.~M.; Hassanpour,~N.; Hook,~D.~W.; Klevansky,~S.~P.; S\"underhauf,~C.;
  Wen,~Z. Behavior of Eigenvalues in a Region of Broken {{PT}} Symmetry.
  \emph{Phys. Rev. A} \textbf{2017}, \emph{95}, 052113\relax
\mciteBstWouldAddEndPuncttrue
\mciteSetBstMidEndSepPunct{\mcitedefaultmidpunct}
{\mcitedefaultendpunct}{\mcitedefaultseppunct}\relax
\EndOfBibitem
\bibitem[Beygi \latin{et~al.}(2018)Beygi, Klevansky, and Bender]{Beygi_2018a}
Beygi,~A.; Klevansky,~S.~P.; Bender,~C.~M. Two- and Four-Dimensional
  Representations of the {{PT}} - and {{CPT}} -Symmetric Fermionic Algebras.
  \emph{Phys. Rev. A} \textbf{2018}, \emph{97}, 032128\relax
\mciteBstWouldAddEndPuncttrue
\mciteSetBstMidEndSepPunct{\mcitedefaultmidpunct}
{\mcitedefaultendpunct}{\mcitedefaultseppunct}\relax
\EndOfBibitem
\bibitem[Liskow \latin{et~al.}(1972)Liskow, Bender, and Schaefer]{Liskow_1972}
Liskow,~D.~H.; Bender,~C.~F.; Schaefer,~H.~F. Bending {{Frequency}} of the
  {{C}} {\textsubscript{3}} {{Molecule}}. \emph{J. Chem. Phys.} \textbf{1972},
  \emph{56}, 5075--5080\relax
\mciteBstWouldAddEndPuncttrue
\mciteSetBstMidEndSepPunct{\mcitedefaultmidpunct}
{\mcitedefaultendpunct}{\mcitedefaultseppunct}\relax
\EndOfBibitem
\bibitem[Peng \latin{et~al.}(2014)Peng, Ozdemir, Rotter, Yilmaz, Liertzer,
  Monifi, Bender, Nori, and Yang]{Peng_2014}
Peng,~B.; Ozdemir,~.~K.; Rotter,~S.; Yilmaz,~H.; Liertzer,~M.; Monifi,~F.;
  Bender,~C.~M.; Nori,~F.; Yang,~L. Loss-Induced Suppression and Revival of
  Lasing. \emph{Science} \textbf{2014}, \emph{346}, 328--332\relax
\mciteBstWouldAddEndPuncttrue
\mciteSetBstMidEndSepPunct{\mcitedefaultmidpunct}
{\mcitedefaultendpunct}{\mcitedefaultseppunct}\relax
\EndOfBibitem
\bibitem[Peng \latin{et~al.}(2014)Peng, \"Ozdemir, Lei, Monifi, Gianfreda,
  Long, Fan, Nori, Bender, and Yang]{Peng_2014a}
Peng,~B.; \"Ozdemir,~{\c S}.~K.; Lei,~F.; Monifi,~F.; Gianfreda,~M.;
  Long,~G.~L.; Fan,~S.; Nori,~F.; Bender,~C.~M.; Yang,~L.
  Parity\textendash{}Time-Symmetric Whispering-Gallery Microcavities.
  \emph{Nat. Phys.} \textbf{2014}, \emph{10}, 394--398\relax
\mciteBstWouldAddEndPuncttrue
\mciteSetBstMidEndSepPunct{\mcitedefaultmidpunct}
{\mcitedefaultendpunct}{\mcitedefaultseppunct}\relax
\EndOfBibitem
\bibitem[Bender(2019)]{BenderPTBook}
Bender,~C.~M. \emph{{{\cal PT}}-Symmetry in Quantum and Classical Physics};
  World Scientific, 2019\relax
\mciteBstWouldAddEndPuncttrue
\mciteSetBstMidEndSepPunct{\mcitedefaultmidpunct}
{\mcitedefaultendpunct}{\mcitedefaultseppunct}\relax
\EndOfBibitem
\bibitem[Dorey \latin{et~al.}()Dorey, Dunning, and Tateo]{Dorey_2001}
Dorey,~P.~E.; Dunning,~C.; Tateo,~R. Spectral equivalences, Bethe Ansatz
  equations, and reality properties in PT-symmetric quantum mechanics. \emph{J.
  Phys. A} \emph{34}, 5679--5704\relax
\mciteBstWouldAddEndPuncttrue
\mciteSetBstMidEndSepPunct{\mcitedefaultmidpunct}
{\mcitedefaultendpunct}{\mcitedefaultseppunct}\relax
\EndOfBibitem
\bibitem[Bittner \latin{et~al.}(2012)Bittner, Dietz, G\"unther, Harney,
  {Miski-Oglu}, Richter, and Sch\"afer]{Bittner_2012}
Bittner,~S.; Dietz,~B.; G\"unther,~U.; Harney,~H.~L.; {Miski-Oglu},~M.;
  Richter,~A.; Sch\"afer,~F. {{PT Symmetry}} and {{Spontaneous Symmetry
  Breaking}} in a {{Microwave Billiard}}. \emph{Phys. Rev. Lett.}
  \textbf{2012}, \emph{108}, 024101\relax
\mciteBstWouldAddEndPuncttrue
\mciteSetBstMidEndSepPunct{\mcitedefaultmidpunct}
{\mcitedefaultendpunct}{\mcitedefaultseppunct}\relax
\EndOfBibitem
\bibitem[Chong \latin{et~al.}(2011)Chong, Ge, and Stone]{Chong_2011}
Chong,~Y.~D.; Ge,~L.; Stone,~A.~D. P {{T}} -{{Symmetry Breaking}} and
  {{Laser}}-{{Absorber Modes}} in {{Optical Scattering Systems}}. \emph{Phys.
  Rev. Lett.} \textbf{2011}, \emph{106}, 093902\relax
\mciteBstWouldAddEndPuncttrue
\mciteSetBstMidEndSepPunct{\mcitedefaultmidpunct}
{\mcitedefaultendpunct}{\mcitedefaultseppunct}\relax
\EndOfBibitem
\bibitem[Chtchelkatchev \latin{et~al.}(2012)Chtchelkatchev, Golubov, Baturina,
  and Vinokur]{Chtchelkatchev_2012}
Chtchelkatchev,~N.~M.; Golubov,~A.~A.; Baturina,~T.~I.; Vinokur,~V.~M.
  Stimulation of the {{Fluctuation Superconductivity}} by {{P T Symmetry}}.
  \emph{Phys. Rev. Lett.} \textbf{2012}, \emph{109}, 150405\relax
\mciteBstWouldAddEndPuncttrue
\mciteSetBstMidEndSepPunct{\mcitedefaultmidpunct}
{\mcitedefaultendpunct}{\mcitedefaultseppunct}\relax
\EndOfBibitem
\bibitem[Doppler \latin{et~al.}(2016)Doppler, Mailybaev, B\"ohm, Kuhl,
  Girschik, Libisch, Milburn, Rabl, Moiseyev, and Rotter]{Doppler_2016}
Doppler,~J.; Mailybaev,~A.~A.; B\"ohm,~J.; Kuhl,~U.; Girschik,~A.; Libisch,~F.;
  Milburn,~T.~J.; Rabl,~P.; Moiseyev,~N.; Rotter,~S. Dynamically Encircling an
  Exceptional Point for Asymmetric Mode Switching. \emph{Nature} \textbf{2016},
  \emph{537}, 76--79\relax
\mciteBstWouldAddEndPuncttrue
\mciteSetBstMidEndSepPunct{\mcitedefaultmidpunct}
{\mcitedefaultendpunct}{\mcitedefaultseppunct}\relax
\EndOfBibitem
\bibitem[Guo \latin{et~al.}(2009)Guo, Salamo, Duchesne, Morandotti,
  {Volatier-Ravat}, Aimez, Siviloglou, and Christodoulides]{Guo_2009}
Guo,~A.; Salamo,~G.~J.; Duchesne,~D.; Morandotti,~R.; {Volatier-Ravat},~M.;
  Aimez,~V.; Siviloglou,~G.~A.; Christodoulides,~D.~N. Observation of {{P T}}
  -{{Symmetry Breaking}} in {{Complex Optical Potentials}}. \emph{Phys. Rev.
  Lett.} \textbf{2009}, \emph{103}, 093902\relax
\mciteBstWouldAddEndPuncttrue
\mciteSetBstMidEndSepPunct{\mcitedefaultmidpunct}
{\mcitedefaultendpunct}{\mcitedefaultseppunct}\relax
\EndOfBibitem
\bibitem[Hang \latin{et~al.}(2013)Hang, Huang, and Konotop]{Hang_2013}
Hang,~C.; Huang,~G.; Konotop,~V.~V. P {{T Symmetry}} with a {{System}} of
  {{Three}}-{{Level Atoms}}. \emph{Phys. Rev. Lett.} \textbf{2013}, \emph{110},
  083604\relax
\mciteBstWouldAddEndPuncttrue
\mciteSetBstMidEndSepPunct{\mcitedefaultmidpunct}
{\mcitedefaultendpunct}{\mcitedefaultseppunct}\relax
\EndOfBibitem
\bibitem[Liertzer \latin{et~al.}(2012)Liertzer, Ge, Cerjan, Stone, T\"ureci,
  and Rotter]{Liertzer_2012}
Liertzer,~M.; Ge,~L.; Cerjan,~A.; Stone,~A.~D.; T\"ureci,~H.~E.; Rotter,~S.
  Pump-{{Induced Exceptional Points}} in {{Lasers}}. \emph{Phys. Rev. Lett.}
  \textbf{2012}, \emph{108}, 173901\relax
\mciteBstWouldAddEndPuncttrue
\mciteSetBstMidEndSepPunct{\mcitedefaultmidpunct}
{\mcitedefaultendpunct}{\mcitedefaultseppunct}\relax
\EndOfBibitem
\bibitem[Longhi(2010)]{Longhi_2010}
Longhi,~S. Optical {{Realization}} of {{Relativistic Non}}-{{Hermitian Quantum
  Mechanics}}. \emph{Phys. Rev. Lett.} \textbf{2010}, \emph{105}, 013903\relax
\mciteBstWouldAddEndPuncttrue
\mciteSetBstMidEndSepPunct{\mcitedefaultmidpunct}
{\mcitedefaultendpunct}{\mcitedefaultseppunct}\relax
\EndOfBibitem
\bibitem[Regensburger \latin{et~al.}(2012)Regensburger, Bersch, Miri,
  Onishchukov, Christodoulides, and Peschel]{Regensburger_2012}
Regensburger,~A.; Bersch,~C.; Miri,~M.-A.; Onishchukov,~G.;
  Christodoulides,~D.~N.; Peschel,~U. Parity\textendash{}Time Synthetic
  Photonic Lattices. \emph{Nature} \textbf{2012}, \emph{488}, 167--171\relax
\mciteBstWouldAddEndPuncttrue
\mciteSetBstMidEndSepPunct{\mcitedefaultmidpunct}
{\mcitedefaultendpunct}{\mcitedefaultseppunct}\relax
\EndOfBibitem
\bibitem[R\"uter \latin{et~al.}(2010)R\"uter, Makris, {El-Ganainy},
  Christodoulides, Segev, and Kip]{Ruter_2010}
R\"uter,~C.~E.; Makris,~K.~G.; {El-Ganainy},~R.; Christodoulides,~D.~N.;
  Segev,~M.; Kip,~D. Observation of Parity\textendash{}Time Symmetry in Optics.
  \emph{Nat. Phys.} \textbf{2010}, \emph{6}, 192--195\relax
\mciteBstWouldAddEndPuncttrue
\mciteSetBstMidEndSepPunct{\mcitedefaultmidpunct}
{\mcitedefaultendpunct}{\mcitedefaultseppunct}\relax
\EndOfBibitem
\bibitem[Schindler \latin{et~al.}(2011)Schindler, Li, Zheng, Ellis, and
  Kottos]{Schindler_2011}
Schindler,~J.; Li,~A.; Zheng,~M.~C.; Ellis,~F.~M.; Kottos,~T. Experimental
  Study of Active {{{\emph{LRC}}}} Circuits with {{PT}} Symmetries. \emph{Phys.
  Rev. A} \textbf{2011}, \emph{84}, 040101\relax
\mciteBstWouldAddEndPuncttrue
\mciteSetBstMidEndSepPunct{\mcitedefaultmidpunct}
{\mcitedefaultendpunct}{\mcitedefaultseppunct}\relax
\EndOfBibitem
\bibitem[Szameit \latin{et~al.}(2011)Szameit, Rechtsman, {Bahat-Treidel}, and
  Segev]{Szameit_2011}
Szameit,~A.; Rechtsman,~M.~C.; {Bahat-Treidel},~O.; Segev,~M. P {{T}} -Symmetry
  in Honeycomb Photonic Lattices. \emph{Phys. Rev. A} \textbf{2011}, \emph{84},
  021806\relax
\mciteBstWouldAddEndPuncttrue
\mciteSetBstMidEndSepPunct{\mcitedefaultmidpunct}
{\mcitedefaultendpunct}{\mcitedefaultseppunct}\relax
\EndOfBibitem
\bibitem[Zhao \latin{et~al.}(2010)Zhao, Schaden, and Wu]{Zhao_2010}
Zhao,~K.~F.; Schaden,~M.; Wu,~Z. Enhanced Magnetic Resonance Signal of
  Spin-Polarized {{Rb}} Atoms near Surfaces of Coated Cells. \emph{Phys. Rev.
  A} \textbf{2010}, \emph{81}, 042903\relax
\mciteBstWouldAddEndPuncttrue
\mciteSetBstMidEndSepPunct{\mcitedefaultmidpunct}
{\mcitedefaultendpunct}{\mcitedefaultseppunct}\relax
\EndOfBibitem
\bibitem[Zheng \latin{et~al.}(2013)Zheng, Hao, and Long]{Zheng_2013}
Zheng,~C.; Hao,~L.; Long,~G.~L. Observation of a Fast Evolution in a
  Parity-Time-Symmetric System. \emph{Philos. Trans. R. Soc. Math. Phys. Eng.
  Sci.} \textbf{2013}, \emph{371}, 20120053--20120053\relax
\mciteBstWouldAddEndPuncttrue
\mciteSetBstMidEndSepPunct{\mcitedefaultmidpunct}
{\mcitedefaultendpunct}{\mcitedefaultseppunct}\relax
\EndOfBibitem
\bibitem[Choi \latin{et~al.}(2018)Choi, Hahn, Yoon, and Song]{Choi_2018}
Choi,~Y.; Hahn,~C.; Yoon,~J.~W.; Song,~S.~H. Observation of an
  Anti-{{PT}}-Symmetric Exceptional Point and Energy-Difference Conserving
  Dynamics in Electrical Circuit Resonators. \emph{Nat. Commun.} \textbf{2018},
  \emph{9}, 2182\relax
\mciteBstWouldAddEndPuncttrue
\mciteSetBstMidEndSepPunct{\mcitedefaultmidpunct}
{\mcitedefaultendpunct}{\mcitedefaultseppunct}\relax
\EndOfBibitem
\bibitem[Goldzak \latin{et~al.}(2018)Goldzak, Mailybaev, and
  Moiseyev]{Goldzak_2018}
Goldzak,~T.; Mailybaev,~A.~A.; Moiseyev,~N. Light Stops at Exceptional Points.
  \emph{Phys. Rev. Lett.} \textbf{2018}, \emph{120}, 013901\relax
\mciteBstWouldAddEndPuncttrue
\mciteSetBstMidEndSepPunct{\mcitedefaultmidpunct}
{\mcitedefaultendpunct}{\mcitedefaultseppunct}\relax
\EndOfBibitem
\bibitem[Heiss and Sannino(1990)Heiss, and Sannino]{Heiss_1990}
Heiss,~W.~D.; Sannino,~A.~L. Avoided Level Crossing and Exceptional Points.
  \emph{J. Phys. Math. Gen.} \textbf{1990}, \emph{23}, 1167--1178\relax
\mciteBstWouldAddEndPuncttrue
\mciteSetBstMidEndSepPunct{\mcitedefaultmidpunct}
{\mcitedefaultendpunct}{\mcitedefaultseppunct}\relax
\EndOfBibitem
\bibitem[Heiss and Sannino(1991)Heiss, and Sannino]{Heiss_1991}
Heiss,~W.~D.; Sannino,~A.~L. Transitional Regions of Finite {{Fermi}} Systems
  and Quantum Chaos. \emph{Phys. Rev. A} \textbf{1991}, \emph{43},
  4159--4166\relax
\mciteBstWouldAddEndPuncttrue
\mciteSetBstMidEndSepPunct{\mcitedefaultmidpunct}
{\mcitedefaultendpunct}{\mcitedefaultseppunct}\relax
\EndOfBibitem
\bibitem[Heiss(1999)]{Heiss_1999}
Heiss,~W.~D. Phases of wave functions and level repulsion. \emph{Eur. Phys. J.
  D} \textbf{1999}, \emph{7}, 1\relax
\mciteBstWouldAddEndPuncttrue
\mciteSetBstMidEndSepPunct{\mcitedefaultmidpunct}
{\mcitedefaultendpunct}{\mcitedefaultseppunct}\relax
\EndOfBibitem
\bibitem[Dorey \latin{et~al.}(2009)Dorey, Dunning, Lishman, and
  Tateo]{Dorey_2009}
Dorey,~P.; Dunning,~C.; Lishman,~A.; Tateo,~R. {{\cal PT}} symmetry breaking
  and exceptional points for a class of inhomogeneous complex potentials.
  \emph{Journal of Physics A: Mathematical and Theoretical} \textbf{2009},
  \emph{42}, 465302\relax
\mciteBstWouldAddEndPuncttrue
\mciteSetBstMidEndSepPunct{\mcitedefaultmidpunct}
{\mcitedefaultendpunct}{\mcitedefaultseppunct}\relax
\EndOfBibitem
\bibitem[Heiss(2012)]{Heiss_2012}
Heiss,~W.~D. The Physics of Exceptional Points. \emph{J. Phys. Math. Theor.}
  \textbf{2012}, \emph{45}, 444016\relax
\mciteBstWouldAddEndPuncttrue
\mciteSetBstMidEndSepPunct{\mcitedefaultmidpunct}
{\mcitedefaultendpunct}{\mcitedefaultseppunct}\relax
\EndOfBibitem
\bibitem[Heiss(2016)]{Heiss_2016}
Heiss,~D. Circling Exceptional Points. \emph{Nat. Phys.} \textbf{2016},
  \emph{12}, 823--824\relax
\mciteBstWouldAddEndPuncttrue
\mciteSetBstMidEndSepPunct{\mcitedefaultmidpunct}
{\mcitedefaultendpunct}{\mcitedefaultseppunct}\relax
\EndOfBibitem
\bibitem[Lefebvre and Moiseyev(2010)Lefebvre, and Moiseyev]{Lefebvre_2010}
Lefebvre,~R.; Moiseyev,~N. Localization of Exceptional Points with {{Pad\'e}}
  Approximants. \emph{J. Phys. B At. Mol. Opt. Phys.} \textbf{2010}, \emph{43},
  095401\relax
\mciteBstWouldAddEndPuncttrue
\mciteSetBstMidEndSepPunct{\mcitedefaultmidpunct}
{\mcitedefaultendpunct}{\mcitedefaultseppunct}\relax
\EndOfBibitem
\bibitem[Mailybaev \latin{et~al.}(2005)Mailybaev, Kirillov, and
  Seyranian]{Mailybaev_2005}
Mailybaev,~A.~A.; Kirillov,~O.~N.; Seyranian,~A.~P. Geometric Phase around
  Exceptional Points. \emph{Phys. Rev. A} \textbf{2005}, \emph{72}\relax
\mciteBstWouldAddEndPuncttrue
\mciteSetBstMidEndSepPunct{\mcitedefaultmidpunct}
{\mcitedefaultendpunct}{\mcitedefaultseppunct}\relax
\EndOfBibitem
\bibitem[Zhang \latin{et~al.}(2018)Zhang, Wang, Hou, and Chan]{Zhang_2018}
Zhang,~X.-L.; Wang,~S.; Hou,~B.; Chan,~C.~T. Dynamically {{Encircling
  Exceptional Points}}: {{{\emph{In}}}}{\emph{ Situ}} {{Control}} of
  {{Encircling Loops}} and the {{Role}} of the {{Starting Point}}. \emph{Phys.
  Rev. X} \textbf{2018}, \emph{8}\relax
\mciteBstWouldAddEndPuncttrue
\mciteSetBstMidEndSepPunct{\mcitedefaultmidpunct}
{\mcitedefaultendpunct}{\mcitedefaultseppunct}\relax
\EndOfBibitem
\bibitem[Yarkony(1996)]{Yarkony_1996}
Yarkony,~D.~R. Diabolical conical intersections. \emph{Rev. Mod. Phys.}
  \textbf{1996}, \emph{68}, 985\relax
\mciteBstWouldAddEndPuncttrue
\mciteSetBstMidEndSepPunct{\mcitedefaultmidpunct}
{\mcitedefaultendpunct}{\mcitedefaultseppunct}\relax
\EndOfBibitem
\bibitem[Jones-Smith and Mathur(2010)Jones-Smith, and Mathur]{Jones-Smith_2010}
Jones-Smith,~K.; Mathur,~H. Non-Hermitian quantum Hamiltonians with
  $\mathcal{P}\mathcal{T}$ symmetry. \emph{Phys. Rev. A} \textbf{2010},
  \emph{82}, 042101\relax
\mciteBstWouldAddEndPuncttrue
\mciteSetBstMidEndSepPunct{\mcitedefaultmidpunct}
{\mcitedefaultendpunct}{\mcitedefaultseppunct}\relax
\EndOfBibitem
\bibitem[Cherbal and Trifonov(2012)Cherbal, and Trifonov]{Cherbal_2012}
Cherbal,~O.; Trifonov,~D.~A. Extended {{PT}} - and {{CPT}} -Symmetric
  Representations of Fermionic Algebras. \emph{Phys. Rev. A} \textbf{2012},
  \emph{85}\relax
\mciteBstWouldAddEndPuncttrue
\mciteSetBstMidEndSepPunct{\mcitedefaultmidpunct}
{\mcitedefaultendpunct}{\mcitedefaultseppunct}\relax
\EndOfBibitem
\bibitem[Beygi and Klevansky(2018)Beygi, and Klevansky]{Beygi_2018b}
Beygi,~A.; Klevansky,~S.~P. No-Signaling Principle and Quantum Brachistochrone
  Problem in {{PT}} -Symmetric Fermionic Two- and Four-Dimensional Models.
  \emph{Phys. Rev. A} \textbf{2018}, \emph{98}, 022105\relax
\mciteBstWouldAddEndPuncttrue
\mciteSetBstMidEndSepPunct{\mcitedefaultmidpunct}
{\mcitedefaultendpunct}{\mcitedefaultseppunct}\relax
\EndOfBibitem
\bibitem[Jones-Smith and Mathur(2014)Jones-Smith, and Mathur]{Jones-Smith_2014}
Jones-Smith,~K.; Mathur,~H. Relativistic non-Hermitian quantum mechanics.
  \emph{Phys. Rev. D} \textbf{2014}, \emph{89}, 125014\relax
\mciteBstWouldAddEndPuncttrue
\mciteSetBstMidEndSepPunct{\mcitedefaultmidpunct}
{\mcitedefaultendpunct}{\mcitedefaultseppunct}\relax
\EndOfBibitem
\bibitem[Ballentine(1998)]{QMModernDevelopment}
Ballentine,~L.~E. \emph{Quantum Mechanics: A Modern Development}; World
  Scientific, 1998\relax
\mciteBstWouldAddEndPuncttrue
\mciteSetBstMidEndSepPunct{\mcitedefaultmidpunct}
{\mcitedefaultendpunct}{\mcitedefaultseppunct}\relax
\EndOfBibitem
\bibitem[Weinberg(1995)]{WeinbergBook}
Weinberg,~S. \emph{The Quantum Theory of Fields}; Cambridge University Press:
  Cambridge, 1995; Vol.~1\relax
\mciteBstWouldAddEndPuncttrue
\mciteSetBstMidEndSepPunct{\mcitedefaultmidpunct}
{\mcitedefaultendpunct}{\mcitedefaultseppunct}\relax
\EndOfBibitem
\bibitem[Zumino(1962)]{Zumino1962}
Zumino,~B. Normal Forms of Complex Matrices. \emph{J. Math. Phys.}
  \textbf{1962}, \emph{3}, 1055\relax
\mciteBstWouldAddEndPuncttrue
\mciteSetBstMidEndSepPunct{\mcitedefaultmidpunct}
{\mcitedefaultendpunct}{\mcitedefaultseppunct}\relax
\EndOfBibitem
\end{mcitethebibliography}

\end{document}